\documentclass[aps,pra,twocolumn,amsmath,amssymb,superscriptaddress,10pt]{revtex4-1}
\usepackage{amsfonts}
\usepackage{amsmath}
\usepackage{amssymb}
\usepackage{graphicx}
\usepackage[T1]{fontenc}
\usepackage[utf8]{inputenc}
\usepackage{hyperref}
\usepackage[dvipsnames]{xcolor}
\usepackage{color}
\usepackage{bm}
\usepackage{comment}
\usepackage{mathtools}
\usepackage[section]{placeins}
\usepackage[capitalize]{cleveref}
\usepackage{footnote}
\usepackage{etoolbox}

\newcommand{\q}[1]{``#1''}
\newcommand{\red}{\textcolor{red}}

\newcommand{\bvm}{\boldsymbol{m}}
\newcommand{\bvr}{\boldsymbol{r}}
\newcommand{\bhr}{\boldsymbol{\hat r}}
\newcommand{\bhx}{\boldsymbol{\hat x}}
\newcommand{\bhy}{\boldsymbol{\hat y}}
\newcommand{\bhz}{\boldsymbol{\hat z}}
\newcommand{\vB}{\boldsymbol{B}}
\newcommand{\vH}{\boldsymbol{H}}
\newcommand{\vM}{\boldsymbol{M}}
\newcommand{\sN}{\mathcal{N}}

\newcommand{\cexp }{\chi_{\text{exp}}}
\newcommand{\cint}{\chi_{\text{int}}}
\newcommand{\cintmc}{\chi_{\text{int}}^{\text{MC}}}
\newcommand{\cexpmc}{\chi_{\text{exp}}^{\text{MC}}}
\newcommand{\Hint}{H_{\text{int}}}
\newcommand{\Hext}{H_{\text{ext}}}
\newcommand{\Bext}{B_{\text{ext}}}
\newcommand{\Hd}{H_{\text{d}}}
\newcommand{\cloc}{\chi_{\text{loc}}}

\newcommand{\bpar}{B_i^{\shortparallel}}
\newcommand{\bpardip}{B_i^{\shortparallel, \text{dip}}}
\newcommand{\bparext}{B_i^{\shortparallel, \text{ext}}}
\newcommand{\bparself}{B_i^{\shortparallel, \text{self}}}
\newcommand{\bhi}{{\bm{\hat{\textnormal{\bfseries\i}}}}}
\newcommand{\bhj}{{\bm{\hat{\textnormal{\bfseries\j}}}}}

\crefrangelabelformat{equation}{(#3#1#4--#5\crefstripprefix{#1}{#2}#6)}

\begin{document}
\title{Microscopic Aspects of Magnetic Lattice Demagnetizing Factors}

\author{M. Twengstr{\"o}m} 
\affiliation{Department of Physics, Royal Institute of Technology, SE-106 91 Stockholm, Sweden}
 \author{L. Bovo} 
 \affiliation{London Centre for Nanotechnology and Department of Physics and Astronomy, University College London, 17-19 Gordon Street, London, WC1H OAH, U.K.}
  \author{M. J. P.  Gingras}
  \affiliation{Department of Physics and  Astronomy, University of Waterloo, Waterloo, Ontario, N2L 3G1, Canada} 
  \affiliation{Canadian Institute for Advanced Research, 180 Dundas St. W., Toronto, Ontario, M5G 1Z8, Canada}
\affiliation{Perimeter Institute for Theoretical Physics, 31 Caroline St. N., Waterloo, Ontario, N2L 2Y5, Canada}
\author{S. T. Bramwell}
\affiliation{London Centre for Nanotechnology and Department of Physics and Astronomy, University College London, 17-19 Gordon Street, London, WC1H OAH, U.K.}
\author{P. Henelius} 
\affiliation{Department of Physics, Royal Institute of Technology, SE-106 91 Stockholm, Sweden}

\begin{abstract}  
The demagnetizing factor $N$ is of both conceptual interest and practical importance. Considering localized magnetic moments on a lattice, we show that for {\it non-ellipsoidal} samples, $N$ depends on the spin dimensionality (Ising, XY, or Heisenberg) and orientation, as well as the sample shape and susceptibility. The generality of this result is demonstrated by means of a recursive analytic calculation as well as detailed Monte Carlo simulations of realistic model spin Hamiltonians. As an important check and application, we also make an accurate experimental determination of $N$ for a representative collective paramagnet (i.e. the Dy$_2$Ti$_2$O$_7$ spin ice compound) and show that the
temperature dependence of the experimentally determined $N$ agrees closely with 
our theoretical calculations. Our conclusion is that the well established practice of approximating the true sample shape with \q{corresponding ellipsoids} for systems with long-range interactions will in many cases overlook important effects stemming from the {\it microscopic} aspects of the system under consideration.
\end{abstract}

\maketitle
\section{Introduction}
Long-range interactions are important in many areas of science, from cosmology, through the gravitational interaction,  to biology, through Coulomb's law. A long-range interaction may be defined in $d$ spatial dimensions by its two-body potential $V(r)$ scaling with distance $r$ as $r^{-\alpha}$ where $\alpha \le d$~\cite{campa09}. The paramount problem in such systems is how to integrate $V(r)$ over an extended system. Following Newton and Euler, the analysis of general systems has been  largely
based on the exact solutions for spheres and ellipsoids~\cite{osborn45,stoner45,beleggia,graef,difratta16}.  This raises the question of whether approximating other shapes to \q{corresponding ellipsoids}~\cite{aharoni98} just neglects uninteresting details or {\it whether there are crucial properties that are lost in the approximation}.
The demagnetizing problem in magnetic systems is a natural setting for exploring this question since it is 
accessible and of intrinsic importance in experiments, and constitutes a paragon for exploring the thermodynamics of long-range interacting systems~\cite{campa09}. Demagnetizing effects are also important in superconductors, while analogues occur, for example, in electric systems~\cite{ponomareva05} (depolarizing factor), in the problem of strain fields around inclusions~\cite{eshelby57}, and in the treatment of avalanching systems in confined geometries~\cite{durin00,csikor07,alava13}.

In an applied magnetic field $H_{\rm ext} = B_{\rm ext}/\mu_0$, the thermodynamic energy of an ellipsoid of volume $V$ and magnetic moment $m$ acquires a contribution $E_{\textrm{mag}}=(\mu_0/2) N m^2/V$, where $N$ is the demagnetizing factor. 
After subtracting $E_{\textrm{mag}}$ from the total energy, differentiation with respect to the magnetization, $M \equiv m/V$, defines the internal field $\Hint \equiv \Hext+\Hd$, where $H_{\rm d} = - NM$ is the demagnetizing field. The {\it intrinsic} magnetic susceptibility $\cint = \partial M/ \partial \Hint$ is a {\it shape-independent material property} derived from the experimentally determined  susceptibility $\cexp \equiv  \partial M/ \partial \Hext$ through
\begin{equation}
\frac{1}{\cint}= \frac{1}{\cexp} - N .
\label{transf}
\end{equation} 

The determination of $N$ is a fundamental problem that dates back to the work of Poisson and Maxwell~\cite{chen91}. In the 1940s, Osborn~\cite{osborn45} and Stoner~\cite{stoner45} tabulated $N$ for general ellipsoids, while more recently,  Aharoni~\cite{aharoni98} treated cuboids in the 
$\cint\!\rightarrow\!0$ limit.  These highly cited  papers bear witness to the importance of
accurately computable and easily accessible demagnetizing factors. Given that {\it i)} it was realized already in the 1920s that $N$ for a non-ellipsoidal sample is a function not only of the sample shape, but also of $\chi_{\text{int}}$ itself~\cite{wuhr23, stab35}, and that {\it ii)} many experiments are routinely performed not on ellipsoids but on cuboids~\cite{quil08, Higa_112}, it is perhaps remarkable 
that it was only very recently that the $\chi$-dependence of $N$ was calculated for cuboids 
away from the $\cint\!\rightarrow\!0$ limit~\cite{chen02,chen05}.

The existence of demagnetizing factors for cuboids suggests that their thermodynamics may be formulated in terms of an internal field, with corrections that become dependent on both shape and temperature~\cite{chen02} (through  $\chi_{\text{int}}$). In this work, we have found that, for magnetic lattices, the demagnetizing factor of cuboids depends {\it also} on the local spin symmetry and allowed orientations of the magnetic moments. With reference to the question posed 
at the very beginning, our result illustrates a case where a long-range interaction integrates in a qualitatively
different way for a cuboid and an ellipsoid, such that the discrete
microscopic nature of the system matters in the former case but not in the latter.  We are aware of only a few previous studies where effects of such  discreteness have been discussed ~\cite{morup83, vedmed02, millev03,ponomareva05}. Our interest in this problem was spurred by the recent experimental observation of anomalous demagnetizing effects in the spin ice material Dy$_2$Ti$_2$O$_7$~\cite{bovo13}.

One may ask whether small differences
in the estimated $N$ really matter for exposing important physics. The answer is found in Eq.~(\ref{transf}).
If $\cexp\ll 1$, then $\cint$ is insensitive to the precise value of $N$. However, in many physical systems that display unusual and interesting magnetic phenomena, $\cexp$ is large, and $\cint$ becomes a sensitive function of $N$. 
Examples include the spin ice materials Dy$_2$Ti$_2$O$_7$ and Ho$_2$Ti$_2$O$_7$, which support magnetic monopole excitations~\cite{Castel08}, and LiHo$_{1-x}$Y$_x$F$_4$ which displays 
ultra-slow relaxation~\cite{bilt12}. Important demagnetizing effects are manifest when an accurately directed field is required:
for example in experiments on the elusive Kasteleyn transition~\cite{fennell07}, sub-lattice pinning~\cite{Ruff2005,Higa_112,Sato1,Sato2} and multiple field-driven transitions~\cite{Schiffer1994};  or else for disentangling the in- and out-of-phase frequency response~\cite{Quilliam2011}. In such cases, quantitative conclusions and accurate tests of theory depend, through $\chi_{\rm int}$, on an accurate knowledge of $N$. Our work illustrates how this may be achieved.

The rest of the paper is organized as follows. In \cref{sec:exp} we discuss how to determine $N$ experimentally. In \cref{sec:iter} we introduce an iterative method for obtaining $N$, and we consider in \cref{sec:mc} a Monte Carlo calculation of $N$. Finally, we close the paper with a discussion in \cref{sec:disc}. For details regarding the experimental and numerical procedures we refer the reader to Appendices \ref{AppSM}-\ref{AppEI}. The effects of short-range interactions are considered in some detail in Appendix~\ref{AppSR}.

\begin{figure}[htb]
    \centering{
    \resizebox{\hsize}{!}{\includegraphics{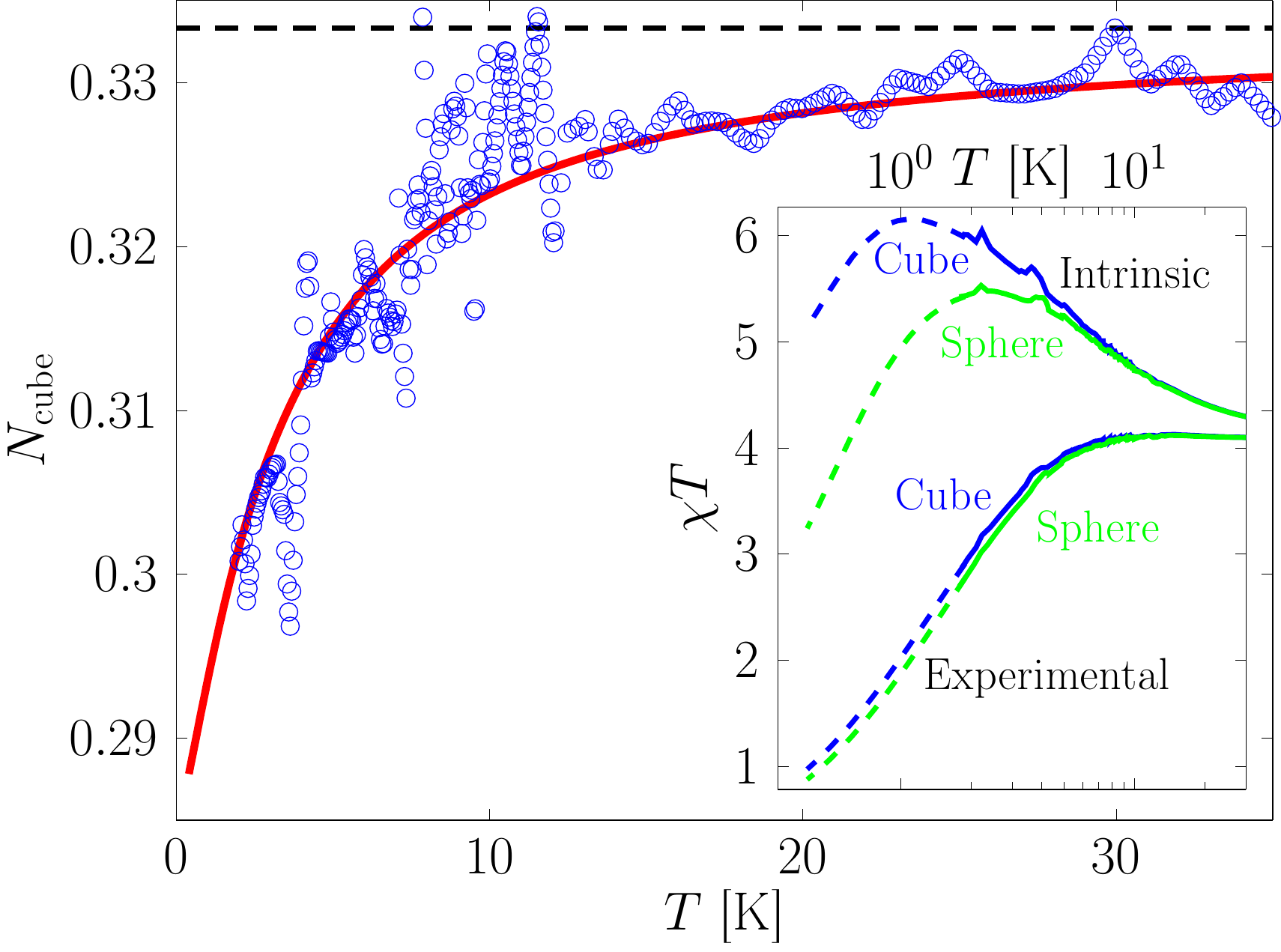}}}
    \caption{Experimentally determined demagnetizing factor for a cube, $N_{\rm cube}$, as a function of temperature, $T$, for Dy$_2$Ti$_2$O$_7$  (blue open circles) compared to our {\it parameter-free} theory (red line). The dashed black line shows the $N_{\rm sphere}=1/3$ exact result~\cite{osborn45}. Inset: The lower solid curves show the susceptibility measured for spherical (green) and cubic samples (blue), from which $N_{\rm cube}$ was derived in this work. The upper curves correspond to data transformed with $N = 1/3$~\cite{aharoni98}, which is incorrect for the cube (upper blue line), but yields the correct  intrinsic susceptibility for the sphere (upper green line). The dashed lines show the predicted theoretical continuation of the experimental data.}
    \label{exp_theory_fit}
\end{figure}

\section{Experimental determination of \texorpdfstring{$N$}{N}}
\label{sec:exp}
To illustrate the importance of the demagnetizing correction, and to test the theory presented below, we first present the {\it experimental} determination of $N$ for a particular case. The localized-moment paramagnet ${\rm Dy_2Ti_2O_7}$ (a spin ice) is well-suited to this purpose as it has a large susceptibility, is crystallographically well-defined (in the cubic space group $Fd \bar{3} m$) with no evidence of crystal distortion~\cite{ruminy16}, and can be accurately cut into high-quality single crystal samples of different shape. Since its spin Hamiltonian has been established in great detail~\cite{Yavo08,hene16}, it is convenient to adopt ${\rm Dy_2Ti_2O_7}$ as a model system for studying the demagnetizing factor.

A sphere of diameter 4 mm and a cube of dimensions $2\times 2\times 2$ mm$^3$, with edges precisely oriented along the cubic crystallographic axes $[100]$, $[010]$ and $[001]$ directions, were commercially hand-cut from different larger crystals of Dy$_2$Ti$_2$O$_7$ provided by D. Prabhakaran~\cite{Prabhak} (see Ref.~\cite{bovo13}). The cube was epi-polished on all sides~\cite{epi}. Crystal shape, orientation, and experimental conditions were carefully controlled to minimize measurement errors; see Appendix~\ref{AppSM}.  The experimental susceptibilities of both the sphere and the cube ($\chi_{\textrm{exp}}^{\textrm{sphere}}$, $\chi_{\textrm{exp}}^{\textrm{cube}}$) were determined from measurements of the magnetic moment. 

Setting the demagnetizing factor of the sphere to $N_{\rm sphere} = 1/3$, that of the cube was determined through Eq.~(\ref{transf}), i.e., $N_{\rm cube}=1/\chi_{\textrm{exp}}^{\textrm{cube}} -1/\chi_{\textrm{exp}}^{\textrm{sphere}} + N_{\rm sphere}$. In order to match the susceptibility of the cube and sphere in the high-$T$ limit, $\chi_{\textrm{exp}}^{\textrm{cube}}$ was shifted by about 1\% ($\chi_{\textrm{exp}}^{\textrm{cube}} \rightarrow \chi_{\textrm{exp}}^{\textrm{cube}}/1.0074$) before calculating $N_{\rm cube}$. Fig.~\ref{exp_theory_fit} shows how the experimental $N_{\rm cube}$ departs significantly from the 1/3 value when $\chi \gtrsim 1$. This is the main experimental result of our study. The inset of the figure compares the uncorrected susceptibility data and the data derived from assuming $N = 1/3$ for both samples. The  predicted theoretical continuation of the experimental data below 2 K (dashed curves) is based on a generalized version of the dipolar spin ice model~\cite{Yavo08,hene16}.

\section{Determination of \texorpdfstring{$N$}{N} via an iterative method}
\label{sec:iter}
In this section we introduce an iterative method to calculate the on-site field distribution inside a linear magnetic material placed in a uniform magnetic field. In the iterative algorithm we first assume that $\Hint$ equals $\Hext$ and calculate the induced local magnetization for an assumed $\cint$. This magnetization generates a demagnetizing field that, in turn, modifies $\Hint$. The resulting field-magnetization equations are iterated until convergence. With the converged field and magnetization distributions in hand, one then computes $N$.

To proceed, we consider a sample of volume $V$ with $\sN$ magnetic moments. As a first case, we focus on Ising moments $\bvm_i=m_i\mu_B{\bhi}$, where ${\bhi}$ is the unit vector in the local Ising direction at site $i$, and $m_i$ is dimensionless. We first determine the component of the local field along the Ising moment at site $i$, $\bpar=\vB_i \cdot {\bhi}$, which is the sum of three contributions:
\begin{equation}
\bpar=\bpardip + \bparext + \bparself , 
\label{field}
\end{equation} 
which we now discuss one by one.

First, the dipolar field at site $i$ produced by {\it all the other} point magnetic dipoles within the sample, 
$\bpardip \equiv \vB_i^{\text{dip}} \cdot {\bhi}$, is given by
the familiar form~\cite{griffiths}
\begin{equation}
\bpardip =
\frac{\mu_0 \mu_{\textrm{B}}}{4\pi}
\sum_{j\ne i} \left(\frac{3 (\bhj\cdot \bhr_{ij})(\bhi \cdot \bhr_{ij}) -\bhj\cdot\bhi}{r_{ij}^3}\right )m_j.
\label{dip}
\end{equation}

Second, we consider an external field in the global $\bhz$ direction, $\vB^{\text{ext}}=B^{\text{ext}}\bhz$, with $\bparext = B^{\text{ext}}\cos\theta_i$, where $\cos\theta_i\equiv \bhz \cdot {\bhi}$, the angle between the direction of the Ising axis  at site $i$ and the direction of $\vB^{\text{ext}}$.

Third, is the contribution from the self-field, $\bparself$. In the classic case of a single point dipole~\cite{griffiths, GriffPaper82}, a term $\frac{2}{3}\mu_0 \mu_{\textrm{B}} \delta(\bvr)$ must be added to ensure that the average magnetic field in a sphere containing the dipole gives the correct macroscopic field. Similarly, we add a self-field to ensure that the internal magnetic field in a uniformly magnetized sample has the expected  value, for example $\vB\!=\!\frac{2}{3}\mu_0\vM$ for a uniformly magnetized sphere or cube~\cite{chen05}. Note that one should, in general, treat the limit of a uniformly magnetized non-ellipsoidal sample with some care. In this work, we are primarily  concerned with paramagnetic samples in the linear response  regime, where a weak magnetic field induces a magnetization proportional to it, as in a typical $\chi$ measurement. For a non-ellipsoidal sample, the induced magnetization is in general  {\it non-uniform}, except in the $\chi\!\rightarrow \!0$ limit. In this limit, $\Hd$ vanishes and, as a consequence, $\Hint$ and $M$ are uniform. Our goal is therefore to determine the self-field so that the magnetic field has the expected value in the $\chi\!\rightarrow \!0$ limit. We demonstrate the basic idea with two examples.

We first take a cubic sample with all moments aligned in the global $\bhz$ direction. In this case $\vB$, $\vM$ and $\vH$ are all aligned with the $\bhz$ direction for which the field equation $\vB\!=\!\mu_0(\vM+\vH)$ reduces to $B^z\!=\!\mu_0(M^z-N_0M^z)\!=\!\frac{2}{3}\mu_0M^z$, where $N_0\!=\!\frac{1}{3}$ is the $\chi\!\rightarrow \!0$ limit of $N$ for a cube~\cite{chen05}. If we consider a simple cubic lattice, it is well known that the lattice sum  vanishes~\cite{jack}. This implies that $B^{z,\text{self}}\!=\!\frac{2}{3}\mu_0M^z$ must be incorporated to ensure the expected net $B^\parallel$ field value.

As a second example, we consider the case of a lattice where all the Ising axes are tilted by the same angle $\theta_i=\theta$ with respect to the $z$-axis, with half the spins tilted to the right and half to the left so that there is no net magnetization in the $\bhx$ or $\bhy$ directions. The total $\vB$, $\vM$, and $\vH$ fields are again in the $\bhz$-direction, but what should the $B^{\shortparallel}$ field parallel to the magnetic moments be? From $\vB\!=\!\mu_0(\vM+\vH)$, it follows that $\vB$, is generated by {\it two terms}, which we discuss separately. We begin with the term generated directly by $\vM$, namely $\vB^{1}\!=\!\mu_0\vM$, or $B^{1,z}\!=\!\mu_0 M^z\!=\!\mu_0M^{\shortparallel}\cos\theta$, where $M^{\shortparallel}$ is the magnetization in the local Ising directions, $M^{\shortparallel}\!=\!V^{-1}\sum_{i=1}^{\sN}\bvm_i\cdot\bhi$. This equation is satisfied by $B^{1,\shortparallel}\!=\!\mu_0M^{\shortparallel}$. The second term, $B^{2,z}\!=\!\mu_0H^z\!=\!-\mu_0N_0M^z\!=\!-\mu_0N_0M^{\shortparallel}\cos\theta$ is generated by $\Hd$.
The field along the magnetic moment is thus $B^{2,\shortparallel}\!=\!-\mu_0N_0M^{\shortparallel}\cos^2\theta$, and the net self-field becomes
\begin{equation}
  \bparself = B^{1,\shortparallel}+B^{2,\shortparallel}= \mu_0\mu_B\frac{\sN}{V}\left[1- N_0\cos^2\theta\right]m_i,
  \label{self}
   \end{equation}
 which is valid when the dipolar lattice sum,  Eq.~(\ref{dip}), vanishes
and  when the average $\vM$ is along $\vB^{\text{ext}}$. For the case in which the lattice sum does not vanish, it must be subtracted from the self-field in order to ensure the expected net field value. 

\cref{field,dip,self} give the local field in terms of the set of local magnetizations, $\{m_i\}$.
With the local fields available we next consider the reverse relation that yields the $\{m_i\}$ induced by $\bpar$. Using $\vM\!=\!\chi\vH$ (linear media), we get $\vB\!=\!\mu_0(\vM+\vM/\chi)\!=\!\mu_0\frac{\chi+1}{\chi}\vM$, leading to
\begin{equation}
m_i=\frac{V}{\sN}\left(\frac{\cloc}{\cloc+1}\right)\frac{\bpar}{\mu_0\mu_B},
\label{magnetization}
\end{equation}
where $\cloc$ is the local susceptibility in the ${\bhi}$ direction, $M^{\shortparallel}=\cloc H^{\shortparallel}$. 

We can now proceed to iterate the expressions for $\bpar$ in Eq.~(\ref{field}) and $m_i$ in Eq.~(\ref{magnetization}) until convergence, and then calculate $N$ from Eq.~(\ref{transf}), where $\cexp$ is given by 
\begin{equation}
\cexp=\cexp^{zz}= \left(\frac{\partial M^z}{\partial H^z_{\text{ext}} }\right)_T =\frac{\mu_0\mu_B}{V B^{\text{ext}}}\sum_{i=1}^{\sN}m_i\cos\theta,
\end{equation}
where we are still considering site independent tilt angles, $\theta_i=\theta$. The intrinsic susceptibility, $\cint$, expresses the relation between $\vB^{\text{ext}}$ and induced $\vM$ under \q{Ewald}, or \q{tin foil}, boundary conditions~\cite{ewal21}, which eliminate demagnetizing fields and correspond to the $N\! = \!0$ limit. As a result, both $\cint$ and $\cloc$ are responses to an {\it internal} field.  While $\cint$ measures the response in the direction of $\Hext$, $\cloc$ measures the response along the local Ising axis ${\bhi}$. With $\bhz \cdot \bhi =\cos\theta$, $H_{\text{ext}}^z\cos\theta$ induces a magnetization $M^{\shortparallel}=\cloc H_{\text{ext}}^z\cos\theta$.
This magnetization, in turn, has a component $M^z=M^{\shortparallel}\cos\theta=\cloc H_{\text{ext}}^z\cos^2\theta$ along $\bhz$, and therefore $\cint =\cloc\cos^2\theta$.

To sum up, once the converged $\bpar$ and $m_i$ distributions have been determined, $N$ is calculated using Eq.~(\ref{transf}),
\begin{equation} 
N =\left[\frac{\mu_0\mu_B}{V B^{\text{ext}}}\sum_{i=1}^{\sN}m_i\cos\theta\right]^{-1} - \frac{1}{\cloc\cos^2\theta}.
\end{equation}

\begin{figure}[htb]
    \centering{
    \resizebox{\hsize}{!}{\includegraphics{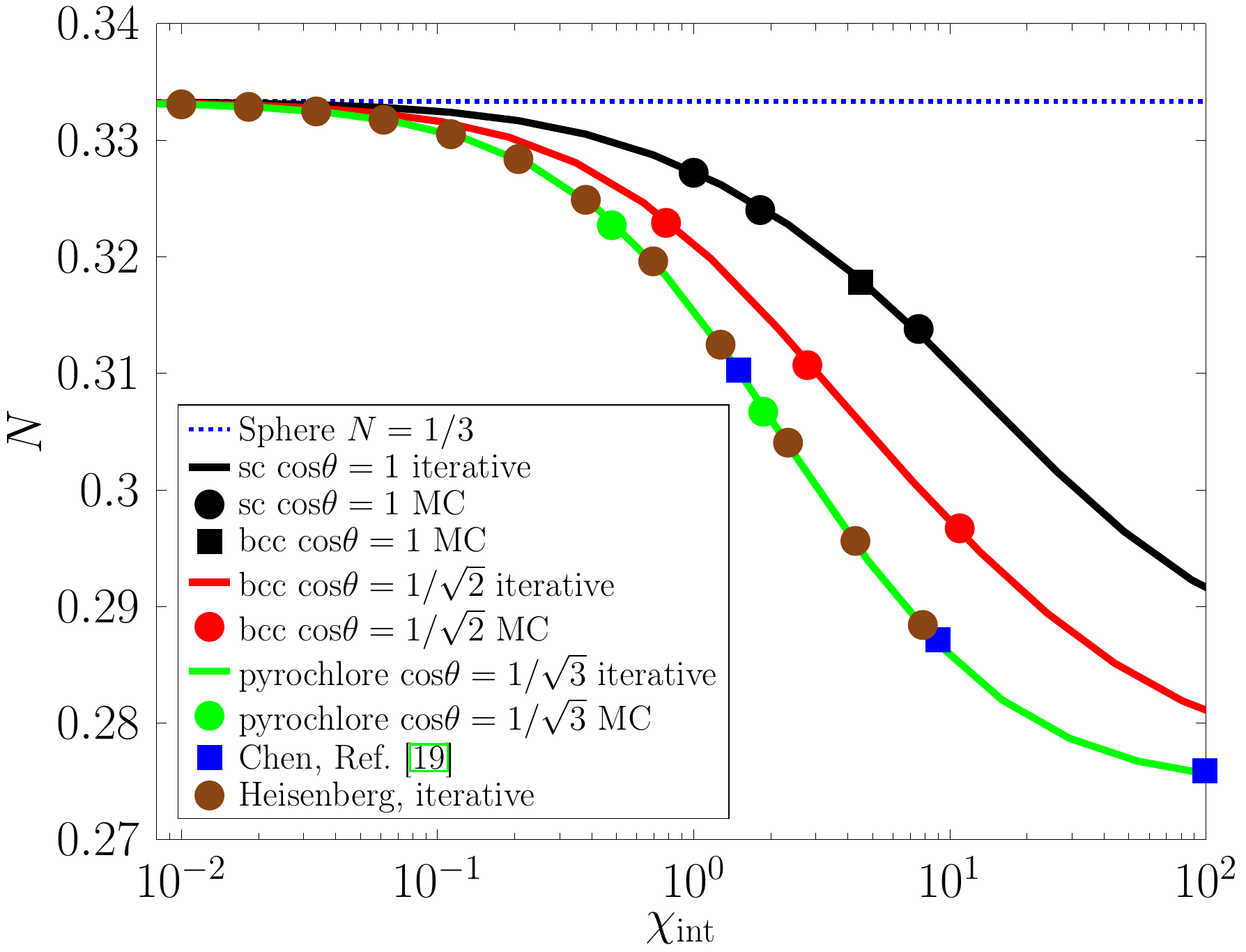}}}
    \caption{$N$ as a function of $\cint$ for cubic samples of various lattices (see main text). Lines show the results from  the iterative calculations, while symbols are Monte Carlo check points. Blue squares are from Chen {\it et al.}~\cite{chen05}. Maroon circles indicate the results of an iterative calculation for isotropic (Heisenberg) spins on an sc lattice. The cosine of the angle $\theta$ between the applied field and the local Ising axes is indicated for each set.} \label{main}
\end{figure}

In Fig.~\ref{exp_theory_fit}, we include $N$ calculated for the pyrochlore lattice using the iterative method (red line), and the main theoretical result is shown in Fig.~\ref{main}, where $N$ is displayed as a function of $\cint$ for cubic samples of the simple cubic (sc) and body centered cubic (bcc) lattices with the Ising direction parallel to $\vB^{\text{ext}}$ $(\cos\theta=1)$. Results (not shown) for a tetragonal lattice, relevant to LiHoF$_4$~\cite{gingras11}, are found to be identical to the sc case. We also display results for a bcc lattice with spins pointing in the $[101]$ and $[\bar 1 0 1 ]$ directions $(\cos\theta=1/\sqrt2)$, and a pyrochlore lattice $(\cos\theta=1/sqrt3)$ built from the conventional cubic unit cell~\cite{melko04}. Finally, we include results for the dipolar model with spherically symmetric Heisenberg spins on an sc lattice. 

\section{Determination of \texorpdfstring{$N$}{N} via Monte Carlo simulations}\label{sec:mc}

With the iterative method, we are able to reach relatively large system sizes of $\mathcal{O}(10^6)$ spins. To verify that this method, which is mean-field like and does not include fluctuations in the $m_i$'s, gives the same result as a full statistical calculation for a given spin Hamiltonian, we have also calculated $N$ using Monte Carlo (MC) simulations for several representative cases (see Fig.~~\ref{main}). For a single data point, the MC approach requires $\mathcal{O}(10^5)$ core hours~\cite{process} to reach the necessary precision for $\mathcal{O}(10^4)$ moments.  Since the iterative formulation contains an internal susceptibility, but no explicit temperature, $T$,  it is necessary to tune either the MC $T$, or the iterative method $\chi_{\text{int}}$,  so that  the MC susceptibility calculated using Ewald boundary conditions, $\chi_{\text{int}}^\text{MC}$, matches the susceptibility from the iterative calculation. We have chosen to adjust the MC temperature, $T$, in order to tune $\cint$ to the desired value. In other words, and to emphasize,  we do not compare a temperature-dependent mean-field theory calculation with a MC calculation at {\it the same nominal temperature}, a calculation which would not generally yield the same $N$ in the thermodynamic limit. For details concerning the numerical methods, we refer the reader to Appendices \ref{AppIM} -  \ref{AppEI}.

For definitiveness, we use the magnetostatic dipolar Hamiltonian
\begin{equation}\label{maghamdip}
\mathcal{H}=\frac{\mu_0\mu^2}{4\pi}\sum_{i > j}\Lambda_{ij}\sigma_i\sigma_j,
\end{equation}
where $\sigma_i=\pm 1$, $\mu$ is the magnetic moment and $\Lambda_{ij}=\left[ (\bhi\cdot\bhj)-3 (\bhi\cdot\bhr_{ij}) (\bhj\cdot\bhr_{ij})\right]/r_{ij}^3,$ and $\chi^{zz}$, in zero field, is determined according to
\begin{equation}
\chi^{zz}=\frac{\partial M^z}{\partial H^z}=\frac{\mu_0\mu^2}{k_{\textup{B}}TV}\left\langle \left(\sum_{i=1}^\sN \sigma_i\cos\theta\right)^2\right\rangle.
\label{sus}
\end{equation}
Using Ewald boundary conditions, we obtain $\cintmc$, while open boundary conditions yield $\cexpmc$, with $N$ obtained from Eq.~(\ref{transf}).

\begin{figure}[htb]
    \centering{
    \resizebox{\hsize}{!}{\includegraphics{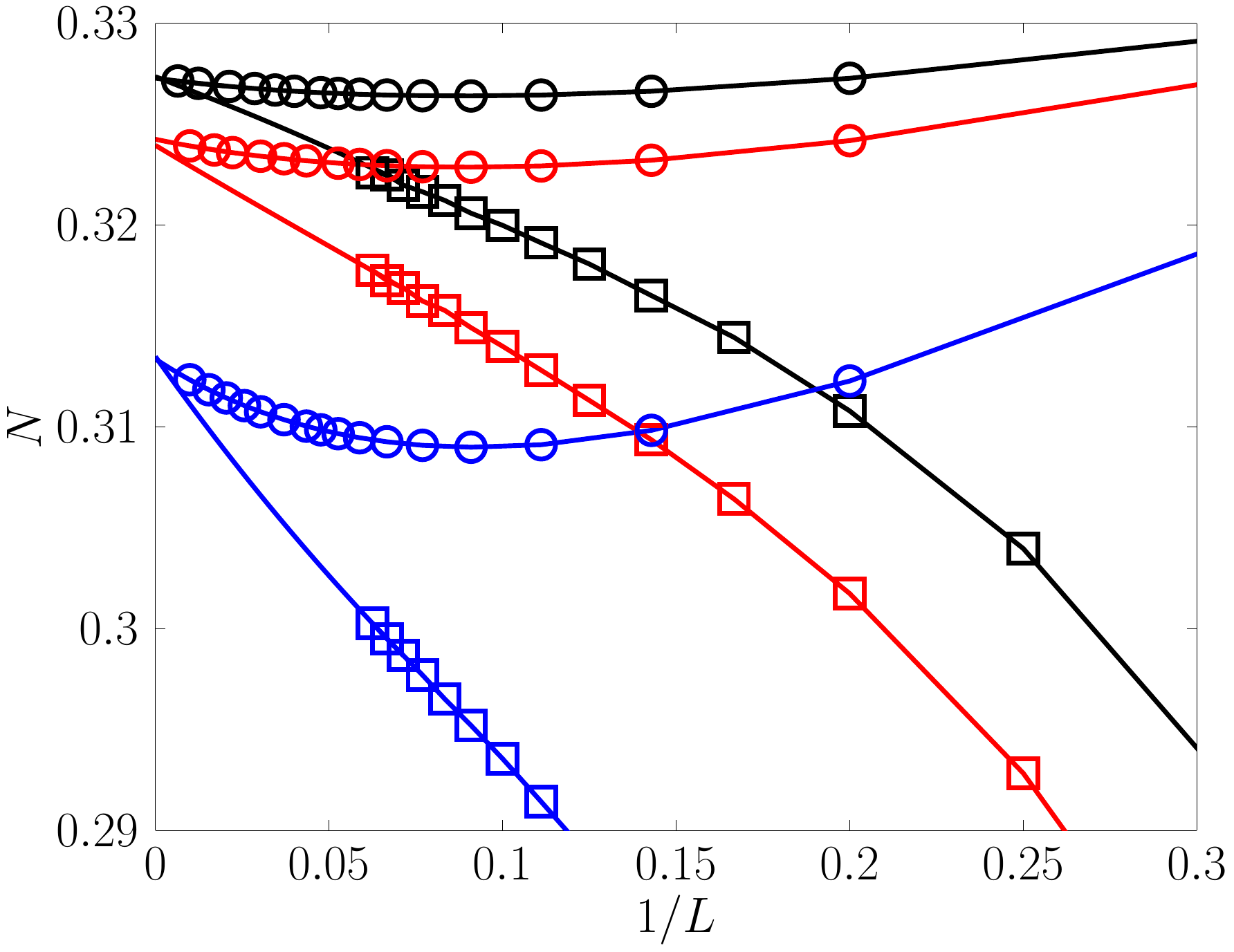}}}
    \caption{$N$ as a function of inverse linear system size, $1/L$, for a cubic sample of an sc lattice with the Ising axes oriented in the $\bhz$ direction. Shown are $\chi_{\text{int}}$=1.00 (black), 1.82 (red) and 7.53 (blue) for the iterative method (circles), and Monte Carlo method (squares). The lines show the extrapolation to the thermodynamic limit using the mathematical expressions described in the main text.} \label{sc}
\end{figure}
Results for the MC method are shown in  Fig.~\ref{main}.
All MC and iterative results have been extrapolated to infinite system size and, in Fig.~\ref{sc}, we compare the system-size dependence of the iterative and MC methods. Results for open boundary conditions are extrapolated using the form $a+b/L+c/L^2$ where the leading $1/L$ term represents a surface to volume ratio effect, while we use  $a+b/L^3+c/L^6$ for periodic boundary conditions, with the leading $1/L^3$-term representing the inverse volume of the  system. These functions yield the best fit to  the data, but we find that the extrapolated value of $N$ is rather insensitive to the precise fitting function, see  Appendix \ref{AppEI}. 

\section{Discussion}
\label{sec:disc}
The key results of this study are threefold. First, we find quantitative agreement between two theoretical methods -- iterative and MC (Fig.~\ref{main}) -- and experiment (Fig.~\ref{exp_theory_fit}), demonstrating that our methods are sound. Second, the explicit $T$ dependence of $N$ for a cuboid has been verified for a real material (Fig.~\ref{exp_theory_fit}). Finally, $N$ is found to depend on the symmetry and direction of the moments (Fig.~\ref{main}). The sc, bcc and LiHoF$_4$ lattices with collinear Ising spins yield the same $N$, indicating that $N$ is not directly sensitive to the lattice. However, turning the local Ising axes away from $\Bext$ causes a more rapid decrease of $N$ with increasing $\cint$. The pyrochlore lattice with tilt angle $\cos\theta=1/\sqrt3$ yields a smaller $N$ than the bcc lattice with $\cos\theta=1/\sqrt2$ for $\cint>0$. The spin ice pyrochlore lattice and the dipolar model with Heisenberg spins yield the same result as the continuum method of Chen {\it et al.}~\cite{chen05}, and we conjecture that models with isotropic $\chi$ will generally follow this behavior~\cite{chen05}. 

Exchange interactions, even when known in detail (e.g., for Dy$_2$Ti$_2$O$_7$~\cite{Yavo08,hene16}),  have not been included in our theoretical models. This is because demagnetizing fields arise solely from the long-range dipolar interactions. The thermodynamic limit for short-range models is well-defined~\cite{levy68,ruelle63,campa14}, and inclusion of short-range interactions does not alter the thermodynamic limit results for $N$; see Appendix \ref{AppSR}. Thermal fluctuations also appear irrelevant in this limit. For ellipsoids, $N$ is calculated from averaged macroscopic fields that do not include thermal fluctuations and, similarly, our mean-field like iterative method captures the essential demagnetizing effects also for cuboids. However, in the non-universal approach to the thermodynamic limit (Fig.~\ref{sc}), there is an expected and significant finite-size difference between the iterative and the MC methods.

What are the experimental implications of our results? If an accurate measurement of $\cint$ is required, then the corrections to $N(\chi\!\rightarrow\!0)$ identified here may be dramatic for $\chi \gtrsim N$. For example, in  the  case of Dy$_2$Ti$_2$O$_7$,  $T \chi_{\textup{int}} (T)$
features a peak, which is easily shifted outside the experimental temperature window by application of the ordinary  $\chi\!=\!0$ demagnetizing correction (see Fig.~\ref{exp_theory_fit} and Ref. ~\cite{bovo13}).
More generally, while the demagnetizing correction is readily controlled for needles or ellipsoids, it is not always easy to prepare real samples with these ideal shapes. This is particularly true of non-metallic and often brittle samples -- e.g., spin ice~\cite{gingras11a} and LiHoF$_4$~\cite{gingras11} -- which have become of significant interest in recent years. Therefore, insofar as cuboidal samples are often the most practical to prepare and control, the best approach may be to use them alongside the theoretical corrections identified in this work.  Our methods are general and  valid for localized-moment magnets independently of details like interaction range and spin dimensionality, and the iterative method can be generalized to non-cuboids. The iterative method could also prove useful for calculating demagnetizing effects in aggregate systems, such as biomedically relevant dispersions of magnetic nanoparticles~\cite{sanchez17}.

In conclusion, considering the demagnetizing problem as a paradigm for the study of long-range interactions, our results confirm that $N$ may be defined for cuboids such that their free energy includes a term $F_{\textrm{mag}} = (\mu_0/2) V N(T) M^2$~\cite{rhodes54} where $M$ is thermodynamically conjugate to $H_{\rm int}$. By going beyond Maxwell's continuum theory, we show that $N$ depends not only on sample shape and $\chi$, but also on microscopic factors: the spin dimensionality and local spin anisotropy. Given that microscopic details affect even such a fundamental and well-studied macroscopic property as $N$, it is interesting to ask how they could affect the thermodynamics of more general long-range interacting systems.

\begin{acknowledgments}
We thank D. Prabhakaran for providing crystals from which the samples were cut, and Tom Fennell and Jeffrey Rau for useful discussions. The simulations were performed on resources provided by the Swedish National Infrastructure for Computing (SNIC) at the Center for High Performance Computing (PDC) at the Royal Institute of Technology (KTH). M.T. and P.H. are supported by the Swedish Research Council (2013-03968), M.T. is grateful for funding from Stiftelsen Olle Engkvist Byggmästare (2014/807), and L.B. is supported by The Leverhulme Trust through the Early Career Fellowship program (ECF2014-284). 
The work at the University of Waterloo was supported by the Canada Research Chair program (M.J.P.G., Tier 1).
This research was supported in part by the Perimeter Institute for Theoretical Physics. Research at the Perimeter Institute is supported by the Government of Canada through Innovation, Science, and Economic Development Canada and by the Province of Ontario through the Ministry of Research, Innovation, and Science. 
\end{acknowledgments}

\appendix
\renewcommand\thefigure{\thesection.\arabic{figure}}
\setcounter{figure}{0}
\renewcommand{\thetable}{\thesection.\arabic{table}} 
\setcounter{table}{0}

\section{Susceptibility measurement}\label{AppSM}
The magnetic susceptibility was measured using a Quantum Design SQUID magnetometer and the crystals were positioned in a cylindrical plastic tube to ensure a uniform magnetic environment. Measurements were performed in the RSO (reciprocating sample option) operating mode to achieve better sensitivity by eliminating low frequency noise. The position of the sample was carefully optimized to minimize misalignment with respect to the applied magnetic field. In particular, the sphere was measured at different positions and orientations in order to confirm the isotropic response and to fully reproduce the results of~\cite{bovo13}. Similarly, the cube, with edges cut along $[001]$, $[010]$ and $[001]$, was measured with the field aligned along all three orientations giving equivalent results, as would be expected.

Different measurements were made on each sample and orientation: low-field susceptibility (at $\mu_0H_0 = 0.0025$, $0.005$ and $0.01$ ${\rm T}$) in field-cooled (FC) versus zero-field-cooled (ZFC) protocol. In addition,  magnetic field sweeps at fixed temperature were performed in order to evaluate the susceptibility accurately and confirm the low-field linear response approximation. The FC versus ZFC susceptibility measurements involved cooling the sample to $1.8$ ${\rm K}$ in zero field, applying the weak magnetic field, measuring the susceptibility while warming up to $350$ ${\rm K}$, cooling to $1.8$ ${\rm K}$ again and finally re-measuring the susceptibility while warming. Before switching the magnetic field off, field scans with small steps were performed in order to estimate the absolute susceptibilities. As expected, and previously reported~\cite{Bramwell00}, no difference between field-cooled and zero-field-cooled magnetization was observed in this temperature range. The magnetization of each sample was averaged over all six repetitions (three fields, two scans each) to minimize the influence of noise.

\section{Iterative method}\label{AppIM}

The  iterative method was implemented using a form of \q{trivial parallelization}, in which the local field at all sites is  calculated in parallel for a given magnetic moment distribution. An MPI allgather call~\cite{mpi} is used in order to achieve good strong scaling~\cite{scaling} when run on many processors, a necessity in order to reach $\mathcal{O}(10^6)$ spins used in this study. The number of iterative steps required to reach convergence increases with increasing susceptibility but is $\mathcal{O}(10^2)$ regardless of the number of spins. Therefore, internode communication is not a bottleneck even though we gather and broadcast a vector equal to the length of the number of spins at every iterative step. A typical run for the largest system sizes ($2\times 10^6$ spins) and 1024 cores~\cite{process} takes around 6 hours and requires roughly 400 communications when the intrinsic susceptibility, $\cint\sim 10$.

\section{Monte Carlo method}\label{AppMC}
The Monte Carlo (MC) method used in this study is mostly based on the Metropolis-Hastings single-spin flip algorithm~\cite{Hast} applied to Ising spins. The exception is a loop algorithm~\cite{loopalgo}, which we applied to the dipolar spin ice Hamiltonian in addition to the single-spin flip algorithm.  

\section{Extrapolation to infinite system size}\label{AppEI}

The  approach to the infinite system size limit of the demagnetizing factor $N$ in the iterative and MC methods is illustrated in  Fig.~3 in the main text. Fig.~3 was generated by selecting three MC temperatures ($16$ K, $10$ K and $3.5$ K), and calculating the susceptibilities extrapolated to infinite system size for these temperatures ($\chi_{\text{int}}^\text{MC}= 1.00, 1.82$ and $7.53$). The iterative method calculations were performed with these susceptibility values for all system sizes, and the MC $T$ was also kept the same for all system sizes.

The functional forms used for the extrapolation also deserve further comments. For the open boundary case, the leading term is of the form $1/L$, the surface to volume ratio. This is numerically confirmed in  Table~\ref{tab_obc}, where the first column gives the fitting function, the second the root-mean-square error (RMSE), and the third column the extrapolated value of $N$. The  MC susceptibility is calculated with open boundary conditions for $\chi_{\text{int}}= 1.82$ (red squares in Fig.~3 in the main text). The smallest RMSE is found in the first and last row of Table~\ref{tab_obc}, both with a leading $1/L$ term. All data for open boundary conditions in this study have been transformed using the form $a+b/L^1+c/L^2$, marked in red in Table~\ref{tab_obc}. 

In Table~\ref{tab_pbc}, the corresponding data for periodic boundary conditions are shown, and we note that the RMSE and extrapolated $N$ are not very sensitive to the precise form of the extrapolation function, but the minimum RMSE is found for the function $a+b/L^3+c/L^6$, which represents an expansion in inverse volume of the surface-free system. All data for periodic boundary conditions in this study have been transformed using the form $a+b/L^3+c/L^6$, marked in red in Table~\ref{tab_pbc}.

\begin{table}[htb]
\begin{tabular}{|c|c|c|}
  \hline
  \hline
  function & RMSE ($10^{-6}$) & N \\
  \hline
  $\red{a+b/L^1+c/L^2}$ &  \red{4.87} & \red{0.3238}\\
  $a+b/L^2+c/L^3$ &  9.63 & 0.3203\\
  $a+b/L^3+c/L^6$ &  83.1 & 0.3177\\
  $a+b/L^3+c/L^4$ &  23.1 & 0.3187\\
  $a+b/L^3+c/L^5$ &  28.9 & 0.3184\\
  $a+b/L^2+c/L^4$ &  13.1 & 0.3198\\
  $a+b/L^1+c/L^3$ &  4.92 & 0.3235\\
  \hline
  \hline
\end{tabular}
\caption{RMS error and extrapolated $N$ for various fitting functions applied to the  MC susceptibility calculated with open boundary conditions for $\chi_{\text{int}}= 1.82$ (red squares in Fig.~3 in the main text). The data points for the seven largest system sizes are included in the fit.}   
\label{tab_obc}
\end{table}

\begin{table}[htb]
\begin{tabular}{|c|c|c|}
  \hline
  \hline
  function & RMSE ($10^{-6}$) & N \\
  \hline
  $a+b/L^1+c/L^2$ &  7.60 & 0.3235\\
  $a+b/L^2+c/L^3$ &  3.87 & 0.3238\\
  $\red{a+b/L^3+c/L^6}$ &  \red{3.68} & \red{0.3238}\\
  $a+b/L^3+c/L^4$ &  3.76 & 0.3238\\
  $a+b/L^3+c/L^5$ &  3.72 & 0.3238\\
  $a+b/L^2+c/L^4$ &  4.02 & 0.3242\\
  $a+b/L^1+c/L^3$ &  3.93 & 0.3238\\
  \hline
  \hline
\end{tabular}
\caption{RMS error and calculated $N$ for various fitting functions applied to the MC susceptibility calculated with periodic boundary conditions for $\chi_{\text{int}}= 1.82$ (red squares in Fig.~3 in the main text). The data points for the seven largest system sizes are included in the fit.}   
\label{tab_pbc}
\end{table}

\section{Short-range exchange interactions}\label{AppSR}
\begin{figure}[htb]
	\centering{
  \resizebox{\hsize}{!}{\includegraphics{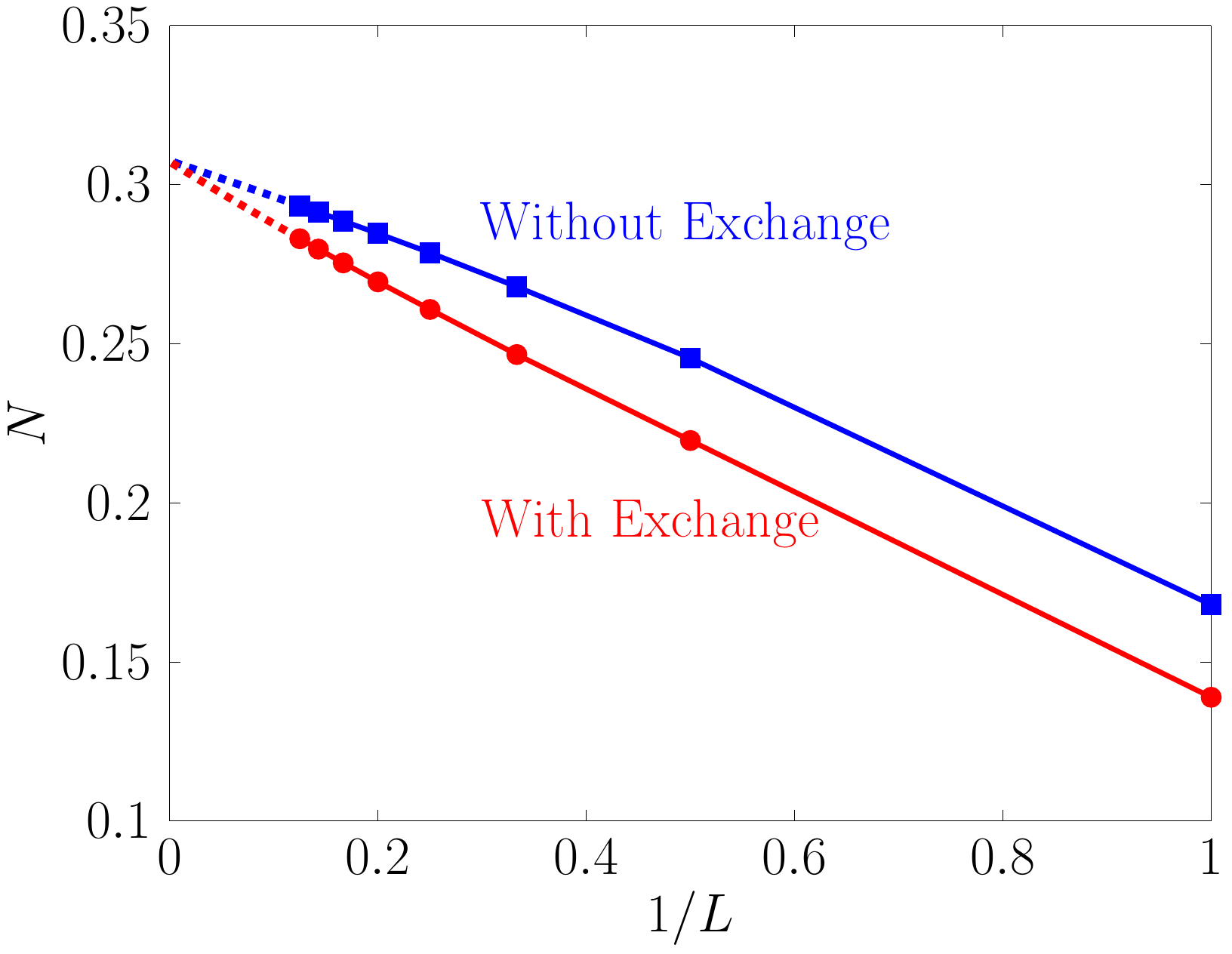}}}
  \caption{Demagnetizing factor as a function of inverse linear system size, $1/L$,  for a dipolar spin ice model containing dipolar terms only and a model containing both dipolar and exchange terms.}
\label{NvsinvL}
\end{figure}

As discussed in the main text, models with short-range interactions have a well defined shape-independent thermodynamic limit and adding exchange interactions to the dipolar Hamiltonian does not alter the demagnetizing factor. We illustrate this explicitly by a numerical MC simulation of the so-called dipolar spin ice model, which has been found to reproduce a number of properties of the Dy$_2$Ti$_2$O$_7$  and Ho$_2$Ti$_2$O$_7$  dipolar spin ice materials~\cite{Hertog2000, Yavo08, hene16}. The  Hamiltonian for this model consists of the dipolar term defined in \cref{maghamdip} and an exchange term of the form
\begin{equation} 
\mathcal{H}_{\textup{exchange}}=\sum_{i>j} J_{ij} \bhi\cdot\bhj\:\sigma_i\sigma_j,
\label{eq:ham} 
\end{equation}
where the strength of the dipolar interaction is given by $D=\mu_0\mu^2/4\pi r_{nn}^3k_B$ with $r_{nn}$ being the nearest-neighbor distance and $k_B$ the Boltzmann's constant. The matrix $J_{ij}$ is the exchange interaction strength between particle $i$ and $j$. Here we consider first ($J_1$), second ($J_2$) and third-nearest-neighbor ($J_{3}$) exchange interactions.

\begin{figure}[htb]
\centering{
  \resizebox{\hsize}{!}{\includegraphics{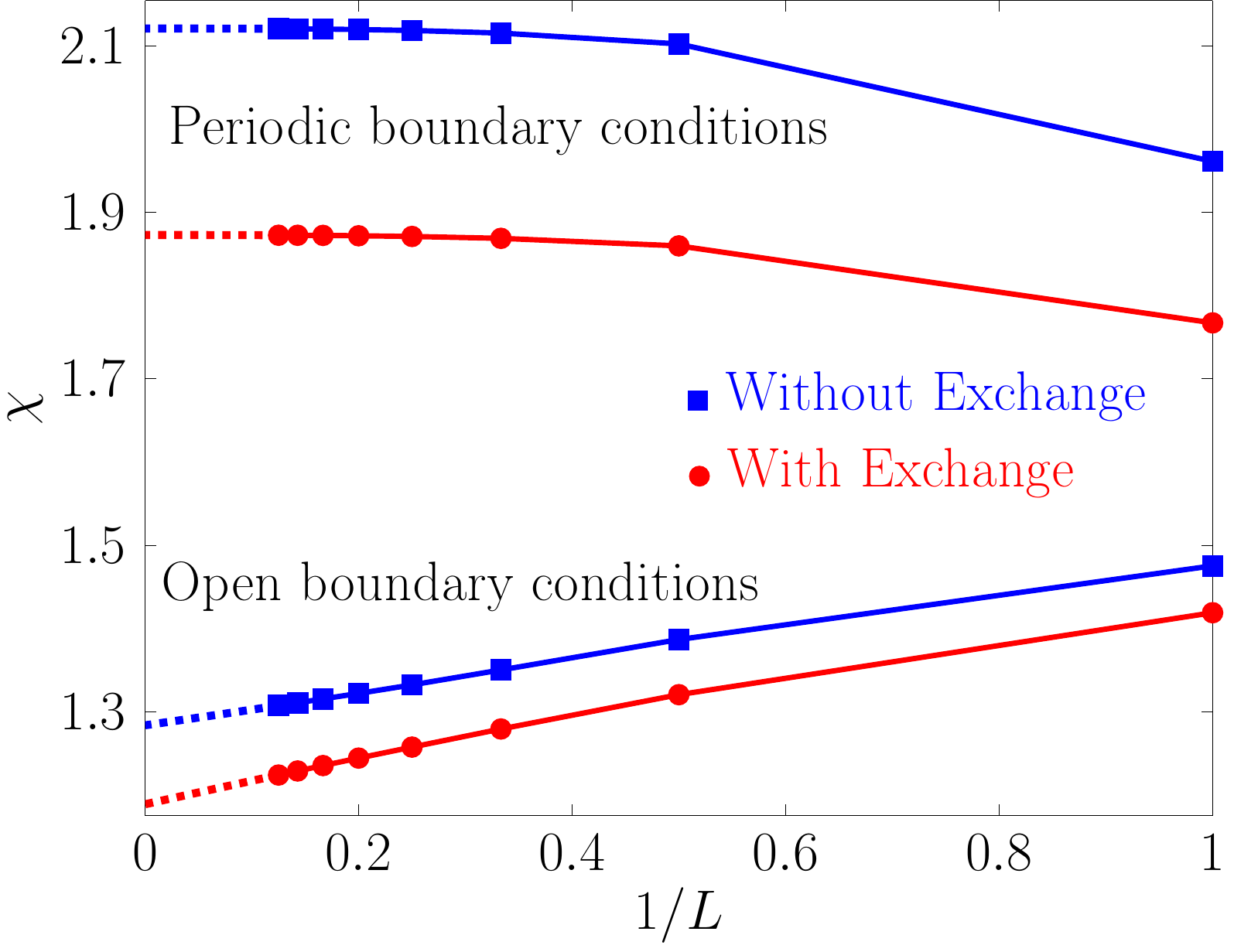}}}
  \caption{Susceptibility as a function of inverse linear system size, $1/L$,  for a model containing dipolar terms only and a model containing both dipolar and exchange terms. For both models the respective results for open and periodic boundary conditions are shown.}
\label{chivsinvL}
\end{figure}

In Fig.~\ref{NvsinvL} we show the demagnetizing factor for this model with parameters ($D=1.3224$ K, $J_1=3.41$ K, $J_2=-0.14$ K, and $J_{3}=0.025$ K, see Ref.~\cite{Yavo08}) and the same model with no exchange interaction ($D=1.3224$ K, $J_1=J_2=J_{3}=0$ K). We expect the infinite system size susceptibility to be dependent on boundary conditions, as shown in Fig.~\ref{chivsinvL}, while the difference of the inverse susceptibilities (demagnetizing factor $N$) is independent of boundary conditions, as shown in Fig.~\ref{NvsinvL}. Hence, we expect no entangling between the dipolar part and the exchange part for the determination of $N$ when both are present.

\begin{thebibliography}{55}%
\makeatletter
\providecommand \@ifxundefined [1]{%
 \@ifx{#1\undefined}
}%
\providecommand \@ifnum [1]{%
 \ifnum #1\expandafter \@firstoftwo
 \else \expandafter \@secondoftwo
 \fi
}%
\providecommand \@ifx [1]{%
 \ifx #1\expandafter \@firstoftwo
 \else \expandafter \@secondoftwo
 \fi
}%
\providecommand \natexlab [1]{#1}%
\providecommand \enquote  [1]{``#1''}%
\providecommand \bibnamefont  [1]{#1}%
\providecommand \bibfnamefont [1]{#1}%
\providecommand \citenamefont [1]{#1}%
\providecommand \href@noop [0]{\@secondoftwo}%
\providecommand \href [0]{\begingroup \@sanitize@url \@href}%
\providecommand \@href[1]{\@@startlink{#1}\@@href}%
\providecommand \@@href[1]{\endgroup#1\@@endlink}%
\providecommand \@sanitize@url [0]{\catcode `\\12\catcode `\$12\catcode
  `\&12\catcode `\#12\catcode `\^12\catcode `\_12\catcode `\%12\relax}%
\providecommand \@@startlink[1]{}%
\providecommand \@@endlink[0]{}%
\providecommand \url  [0]{\begingroup\@sanitize@url \@url }%
\providecommand \@url [1]{\endgroup\@href {#1}{\urlprefix }}%
\providecommand \urlprefix  [0]{URL }%
\providecommand \Eprint [0]{\href }%
\providecommand \doibase [0]{http://dx.doi.org/}%
\providecommand \selectlanguage [0]{\@gobble}%
\providecommand \bibinfo  [0]{\@secondoftwo}%
\providecommand \bibfield  [0]{\@secondoftwo}%
\providecommand \translation [1]{[#1]}%
\providecommand \BibitemOpen [0]{}%
\providecommand \bibitemStop [0]{}%
\providecommand \bibitemNoStop [0]{.\EOS\space}%
\providecommand \EOS [0]{\spacefactor3000\relax}%
\providecommand \BibitemShut  [1]{\csname bibitem#1\endcsname}%
\let\auto@bib@innerbib\@empty
\bibitem [{\citenamefont {Campa}\ \emph {et~al.}(2009)\citenamefont {Campa},
  \citenamefont {Dauxois},\ and\ \citenamefont {Ruffo}}]{campa09}%
  \BibitemOpen
  \bibfield  {author} {\bibinfo {author} {\bibfnamefont {A.}~\bibnamefont
  {Campa}}, \bibinfo {author} {\bibfnamefont {T.}~\bibnamefont {Dauxois}}, \
  and\ \bibinfo {author} {\bibfnamefont {S.}~\bibnamefont {Ruffo}},\
  }\href@noop {} {\bibfield  {journal} {\bibinfo  {journal} {Phys. Rep.}\
  }\textbf {\bibinfo {volume} {480}},\ \bibinfo {pages} {57} (\bibinfo {year}
  {2009})}\BibitemShut {NoStop}%
\bibitem [{\citenamefont {Osborn}(1945)}]{osborn45}%
  \BibitemOpen
  \bibfield  {author} {\bibinfo {author} {\bibfnamefont {J.~A.}\ \bibnamefont
  {Osborn}},\ }\href@noop {} {\bibfield  {journal} {\bibinfo  {journal} {Phys.
  Rev.}\ }\textbf {\bibinfo {volume} {67}},\ \bibinfo {pages} {351} (\bibinfo
  {year} {1945})}\BibitemShut {NoStop}%
\bibitem [{\citenamefont {Stoner}(1945)}]{stoner45}%
  \BibitemOpen
  \bibfield  {author} {\bibinfo {author} {\bibfnamefont {E.~C.}\ \bibnamefont
  {Stoner}},\ }\href@noop {} {\bibfield  {journal} {\bibinfo  {journal}
  {Philos. Mag.}\ }\textbf {\bibinfo {volume} {36}},\ \bibinfo {pages} {803}
  (\bibinfo {year} {1945})}\BibitemShut {NoStop}%
\bibitem [{\citenamefont {Beleggia}\ \emph
  {et~al.}(2006{\natexlab{a}})\citenamefont {Beleggia}, \citenamefont
  {De~Graef},\ and\ \citenamefont {Millev}}]{beleggia}%
  \BibitemOpen
  \bibfield  {author} {\bibinfo {author} {\bibfnamefont {M.}~\bibnamefont
  {Beleggia}}, \bibinfo {author} {\bibfnamefont {M.}~\bibnamefont {De~Graef}},
  \ and\ \bibinfo {author} {\bibfnamefont {Y.~T.}\ \bibnamefont {Millev}},\
  }\href@noop {} {\bibfield  {journal} {\bibinfo  {journal} {Philos. Mag.}\
  }\textbf {\bibinfo {volume} {86}},\ \bibinfo {pages} {2451} (\bibinfo {year}
  {2006}{\natexlab{a}})}\BibitemShut {NoStop}%
\bibitem [{\citenamefont {Beleggia}\ \emph
  {et~al.}(2006{\natexlab{b}})\citenamefont {Beleggia}, \citenamefont
  {De~Graef},\ and\ \citenamefont {Millev}}]{graef}%
  \BibitemOpen
  \bibfield  {author} {\bibinfo {author} {\bibfnamefont {M.}~\bibnamefont
  {Beleggia}}, \bibinfo {author} {\bibfnamefont {M.}~\bibnamefont {De~Graef}},
  \ and\ \bibinfo {author} {\bibfnamefont {Y.~T.}\ \bibnamefont {Millev}},\
  }\href@noop {} {\bibfield  {journal} {\bibinfo  {journal} {J. Phys. D}\
  }\textbf {\bibinfo {volume} {39}},\ \bibinfo {pages} {891} (\bibinfo {year}
  {2006}{\natexlab{b}})}\BibitemShut {NoStop}%
\bibitem [{\citenamefont {Di~Fratta}(2016)}]{difratta16}%
  \BibitemOpen
  \bibfield  {author} {\bibinfo {author} {\bibfnamefont {G.}~\bibnamefont
  {Di~Fratta}},\ }\href@noop {} {\bibfield  {journal} {\bibinfo  {journal}
  {Proc. R. Soc. A}\ }\textbf {\bibinfo {volume} {472}},\ \bibinfo {pages}
  {0197} (\bibinfo {year} {2016})}\BibitemShut {NoStop}%
\bibitem [{\citenamefont {Aharoni}(1998)}]{aharoni98}%
  \BibitemOpen
  \bibfield  {author} {\bibinfo {author} {\bibfnamefont {A.}~\bibnamefont
  {Aharoni}},\ }\href@noop {} {\bibfield  {journal} {\bibinfo  {journal} {J.
  Appl. Phys.}\ }\textbf {\bibinfo {volume} {83}},\ \bibinfo {pages} {3432}
  (\bibinfo {year} {1998})}\BibitemShut {NoStop}%
\bibitem [{\citenamefont {Ponomareva}\ \emph {et~al.}(2005)\citenamefont
  {Ponomareva}, \citenamefont {Naumov}, \citenamefont {Kornev}, \citenamefont
  {Fu},\ and\ \citenamefont {Bellaiche}}]{ponomareva05}%
  \BibitemOpen
  \bibfield  {author} {\bibinfo {author} {\bibfnamefont {I.}~\bibnamefont
  {Ponomareva}}, \bibinfo {author} {\bibfnamefont {I.~I.}\ \bibnamefont
  {Naumov}}, \bibinfo {author} {\bibfnamefont {I.}~\bibnamefont {Kornev}},
  \bibinfo {author} {\bibfnamefont {H.}~\bibnamefont {Fu}}, \ and\ \bibinfo
  {author} {\bibfnamefont {L.}~\bibnamefont {Bellaiche}},\ }\href@noop {}
  {\bibfield  {journal} {\bibinfo  {journal} {Phys. Rev. B}\ }\textbf {\bibinfo
  {volume} {72}},\ \bibinfo {pages} {140102} (\bibinfo {year}
  {2005})}\BibitemShut {NoStop}%
\bibitem [{\citenamefont {Eshelby}(1957)}]{eshelby57}%
  \BibitemOpen
  \bibfield  {author} {\bibinfo {author} {\bibfnamefont {J.~D.}\ \bibnamefont
  {Eshelby}},\ }\href@noop {} {\bibfield  {journal} {\bibinfo  {journal} {Proc.
  R. Soc. A}\ }\textbf {\bibinfo {volume} {241}},\ \bibinfo {pages} {376}
  (\bibinfo {year} {1957})}\BibitemShut {NoStop}%
\bibitem [{\citenamefont {Durin}\ and\ \citenamefont
  {Zapperi}(2000)}]{durin00}%
  \BibitemOpen
  \bibfield  {author} {\bibinfo {author} {\bibfnamefont {G.}~\bibnamefont
  {Durin}}\ and\ \bibinfo {author} {\bibfnamefont {S.}~\bibnamefont
  {Zapperi}},\ }\href@noop {} {\bibfield  {journal} {\bibinfo  {journal} {J.
  Appl. Phys.}\ }\textbf {\bibinfo {volume} {87}},\ \bibinfo {pages} {7031}
  (\bibinfo {year} {2000})}\BibitemShut {NoStop}%
\bibitem [{\citenamefont {Csikor}\ \emph {et~al.}(2007)\citenamefont {Csikor},
  \citenamefont {Motz}, \citenamefont {Weygand}, \citenamefont {Zaiser},\ and\
  \citenamefont {Zapperi}}]{csikor07}%
  \BibitemOpen
  \bibfield  {author} {\bibinfo {author} {\bibfnamefont {F.~F.}\ \bibnamefont
  {Csikor}}, \bibinfo {author} {\bibfnamefont {C.}~\bibnamefont {Motz}},
  \bibinfo {author} {\bibfnamefont {D.}~\bibnamefont {Weygand}}, \bibinfo
  {author} {\bibfnamefont {M.}~\bibnamefont {Zaiser}}, \ and\ \bibinfo {author}
  {\bibfnamefont {S.}~\bibnamefont {Zapperi}},\ }\href@noop {} {\bibfield
  {journal} {\bibinfo  {journal} {Science}\ }\textbf {\bibinfo {volume}
  {318}},\ \bibinfo {pages} {251} (\bibinfo {year} {2007})}\BibitemShut
  {NoStop}%
\bibitem [{\citenamefont {Laurson}\ \emph {et~al.}(2013)\citenamefont
  {Laurson}, \citenamefont {Illa}, \citenamefont {Santucci}, \citenamefont
  {Tallakstad}, \citenamefont {M{\aa}l{\o}y},\ and\ \citenamefont
  {Alava}}]{alava13}%
  \BibitemOpen
  \bibfield  {author} {\bibinfo {author} {\bibfnamefont {L.}~\bibnamefont
  {Laurson}}, \bibinfo {author} {\bibfnamefont {X.}~\bibnamefont {Illa}},
  \bibinfo {author} {\bibfnamefont {S.}~\bibnamefont {Santucci}}, \bibinfo
  {author} {\bibfnamefont {K.-T.}\ \bibnamefont {Tallakstad}}, \bibinfo
  {author} {\bibfnamefont {K.~J.}\ \bibnamefont {M{\aa}l{\o}y}}, \ and\
  \bibinfo {author} {\bibfnamefont {M.~J.}\ \bibnamefont {Alava}},\ }\href@noop
  {} {\bibfield  {journal} {\bibinfo  {journal} {Nat. Comm.}\ }\textbf
  {\bibinfo {volume} {4}},\ \bibinfo {pages} {2927} (\bibinfo {year}
  {2013})}\BibitemShut {NoStop}%
\bibitem [{\citenamefont {Chen}\ \emph {et~al.}(1991)\citenamefont {Chen},
  \citenamefont {Brug},\ and\ \citenamefont {Goldfarb}}]{chen91}%
  \BibitemOpen
  \bibfield  {author} {\bibinfo {author} {\bibfnamefont {D.-X.}\ \bibnamefont
  {Chen}}, \bibinfo {author} {\bibfnamefont {J.~A.}\ \bibnamefont {Brug}}, \
  and\ \bibinfo {author} {\bibfnamefont {R.~B.}\ \bibnamefont {Goldfarb}},\
  }\href@noop {} {\bibfield  {journal} {\bibinfo  {journal} {IEEE Trans.
  Magn.}\ }\textbf {\bibinfo {volume} {27}},\ \bibinfo {pages} {3601} (\bibinfo
  {year} {1991})}\BibitemShut {NoStop}%
\bibitem [{\citenamefont {W\"urschmidt}(1923)}]{wuhr23}%
  \BibitemOpen
  \bibfield  {author} {\bibinfo {author} {\bibfnamefont {J.}~\bibnamefont
  {W\"urschmidt}},\ }\href@noop {} {\bibfield  {journal} {\bibinfo  {journal}
  {Z. Phys.}\ }\textbf {\bibinfo {volume} {12}},\ \bibinfo {pages} {128}
  (\bibinfo {year} {1923})}\BibitemShut {NoStop}%
\bibitem [{\citenamefont {St\"ablein}\ and\ \citenamefont
  {Schlechtweg}(1935)}]{stab35}%
  \BibitemOpen
  \bibfield  {author} {\bibinfo {author} {\bibfnamefont {F.}~\bibnamefont
  {St\"ablein}}\ and\ \bibinfo {author} {\bibfnamefont {H.}~\bibnamefont
  {Schlechtweg}},\ }\href@noop {} {\bibfield  {journal} {\bibinfo  {journal}
  {Z. Phys.}\ }\textbf {\bibinfo {volume} {95}},\ \bibinfo {pages} {630}
  (\bibinfo {year} {1935})}\BibitemShut {NoStop}%
\bibitem [{\citenamefont {Quilliam}\ \emph {et~al.}(2008)\citenamefont
  {Quilliam}, \citenamefont {Meng}, \citenamefont {Mugford},\ and\
  \citenamefont {Kycia}}]{quil08}%
  \BibitemOpen
  \bibfield  {author} {\bibinfo {author} {\bibfnamefont {J.~A.}\ \bibnamefont
  {Quilliam}}, \bibinfo {author} {\bibfnamefont {S.}~\bibnamefont {Meng}},
  \bibinfo {author} {\bibfnamefont {C.~G.~A.}\ \bibnamefont {Mugford}}, \ and\
  \bibinfo {author} {\bibfnamefont {J.~B.}\ \bibnamefont {Kycia}},\ }\href@noop
  {} {\bibfield  {journal} {\bibinfo  {journal} {Phys. Rev. Lett.}\ }\textbf
  {\bibinfo {volume} {101}},\ \bibinfo {pages} {187204} (\bibinfo {year}
  {2008})}\BibitemShut {NoStop}%
\bibitem [{\citenamefont {Higashinaka}\ and\ \citenamefont
  {Maeno}(2005)}]{Higa_112}%
  \BibitemOpen
  \bibfield  {author} {\bibinfo {author} {\bibfnamefont {R.}~\bibnamefont
  {Higashinaka}}\ and\ \bibinfo {author} {\bibfnamefont {Y.}~\bibnamefont
  {Maeno}},\ }\href@noop {} {\bibfield  {journal} {\bibinfo  {journal} {Phys.
  Rev. Lett.}\ }\textbf {\bibinfo {volume} {95}},\ \bibinfo {pages} {237208}
  (\bibinfo {year} {2005})}\BibitemShut {NoStop}%
\bibitem [{\citenamefont {Chen}\ \emph {et~al.}(2002)\citenamefont {Chen},
  \citenamefont {Pardo},\ and\ \citenamefont {Sanchez}}]{chen02}%
  \BibitemOpen
  \bibfield  {author} {\bibinfo {author} {\bibfnamefont {D.-X.}\ \bibnamefont
  {Chen}}, \bibinfo {author} {\bibfnamefont {E.}~\bibnamefont {Pardo}}, \ and\
  \bibinfo {author} {\bibfnamefont {A.}~\bibnamefont {Sanchez}},\ }\href@noop
  {} {\bibfield  {journal} {\bibinfo  {journal} {IEEE Trans. Magn.}\ }\textbf
  {\bibinfo {volume} {38}},\ \bibinfo {pages} {1742} (\bibinfo {year}
  {2002})}\BibitemShut {NoStop}%
\bibitem [{\citenamefont {Chen}\ \emph {et~al.}(2005)\citenamefont {Chen},
  \citenamefont {Pardo},\ and\ \citenamefont {Sanchez}}]{chen05}%
  \BibitemOpen
  \bibfield  {author} {\bibinfo {author} {\bibfnamefont {D.-X.}\ \bibnamefont
  {Chen}}, \bibinfo {author} {\bibfnamefont {E.}~\bibnamefont {Pardo}}, \ and\
  \bibinfo {author} {\bibfnamefont {A.}~\bibnamefont {Sanchez}},\ }\href@noop
  {} {\bibfield  {journal} {\bibinfo  {journal} {IEEE Trans. Magn.}\ }\textbf
  {\bibinfo {volume} {41}},\ \bibinfo {pages} {2077} (\bibinfo {year}
  {2005})}\BibitemShut {NoStop}%
\bibitem [{\citenamefont {Christensen}\ and\ \citenamefont
  {M{\o}rup}(1983)}]{morup83}%
  \BibitemOpen
  \bibfield  {author} {\bibinfo {author} {\bibfnamefont {P.~H.}\ \bibnamefont
  {Christensen}}\ and\ \bibinfo {author} {\bibfnamefont {S.}~\bibnamefont
  {M{\o}rup}},\ }\href@noop {} {\bibfield  {journal} {\bibinfo  {journal} {J.
  Magn. Magn. Mater.}\ }\textbf {\bibinfo {volume} {35}},\ \bibinfo {pages}
  {130} (\bibinfo {year} {1983})}\BibitemShut {NoStop}%
\bibitem [{\citenamefont {Vedmedenko}\ \emph {et~al.}(2003)\citenamefont
  {Vedmedenko}, \citenamefont {Oepen},\ and\ \citenamefont
  {Kirschner}}]{vedmed02}%
  \BibitemOpen
  \bibfield  {author} {\bibinfo {author} {\bibfnamefont {E.~Y.}\ \bibnamefont
  {Vedmedenko}}, \bibinfo {author} {\bibfnamefont {H.~P.}\ \bibnamefont
  {Oepen}}, \ and\ \bibinfo {author} {\bibfnamefont {J.}~\bibnamefont
  {Kirschner}},\ }\href@noop {} {\bibfield  {journal} {\bibinfo  {journal} {J.
  Magn. Magn. Mater.}\ }\textbf {\bibinfo {volume} {256}},\ \bibinfo {pages}
  {237} (\bibinfo {year} {2003})}\BibitemShut {NoStop}%
\bibitem [{\citenamefont {Millev}\ \emph {et~al.}(2003)\citenamefont {Millev},
  \citenamefont {Vedmedenko},\ and\ \citenamefont {Oepen}}]{millev03}%
  \BibitemOpen
  \bibfield  {author} {\bibinfo {author} {\bibfnamefont {Y.~T.}\ \bibnamefont
  {Millev}}, \bibinfo {author} {\bibfnamefont {E.}~\bibnamefont {Vedmedenko}},
  \ and\ \bibinfo {author} {\bibfnamefont {H.~P.}\ \bibnamefont {Oepen}},\
  }\href@noop {} {\bibfield  {journal} {\bibinfo  {journal} {J. Phys. D}\
  }\textbf {\bibinfo {volume} {36}},\ \bibinfo {pages} {2945} (\bibinfo {year}
  {2003})}\BibitemShut {NoStop}%
\bibitem [{\citenamefont {Bovo}\ \emph {et~al.}(2013)\citenamefont {Bovo},
  \citenamefont {Jaubert}, \citenamefont {Holdsworth},\ and\ \citenamefont
  {Bramwell}}]{bovo13}%
  \BibitemOpen
  \bibfield  {author} {\bibinfo {author} {\bibfnamefont {L.}~\bibnamefont
  {Bovo}}, \bibinfo {author} {\bibfnamefont {L.~D.~C.}\ \bibnamefont
  {Jaubert}}, \bibinfo {author} {\bibfnamefont {P.~C.~W.}\ \bibnamefont
  {Holdsworth}}, \ and\ \bibinfo {author} {\bibfnamefont {S.~T.}\ \bibnamefont
  {Bramwell}},\ }\href@noop {} {\bibfield  {journal} {\bibinfo  {journal} {J.
  Phys.: Condens. Matter}\ }\textbf {\bibinfo {volume} {25}},\ \bibinfo {pages}
  {386002} (\bibinfo {year} {2013})}\BibitemShut {NoStop}%
\bibitem [{\citenamefont {Castelnovo}\ \emph {et~al.}(2008)\citenamefont
  {Castelnovo}, \citenamefont {Moessner},\ and\ \citenamefont
  {Sondhi}}]{Castel08}%
  \BibitemOpen
  \bibfield  {author} {\bibinfo {author} {\bibfnamefont {C.}~\bibnamefont
  {Castelnovo}}, \bibinfo {author} {\bibfnamefont {R.}~\bibnamefont
  {Moessner}}, \ and\ \bibinfo {author} {\bibfnamefont {S.~L.}\ \bibnamefont
  {Sondhi}},\ }\href@noop {} {\bibfield  {journal} {\bibinfo  {journal}
  {Nature}\ }\textbf {\bibinfo {volume} {451}},\ \bibinfo {pages} {42}
  (\bibinfo {year} {2008})}\BibitemShut {NoStop}%
\bibitem [{\citenamefont {Biltmo}\ and\ \citenamefont
  {Henelius}(2012)}]{bilt12}%
  \BibitemOpen
  \bibfield  {author} {\bibinfo {author} {\bibfnamefont {A.}~\bibnamefont
  {Biltmo}}\ and\ \bibinfo {author} {\bibfnamefont {P.}~\bibnamefont
  {Henelius}},\ }\href@noop {} {\bibfield  {journal} {\bibinfo  {journal} {Nat.
  Comm.}\ }\textbf {\bibinfo {volume} {3}},\ \bibinfo {pages} {857} (\bibinfo
  {year} {2012})}\BibitemShut {NoStop}%
\bibitem [{\citenamefont {Fennell}\ \emph {et~al.}(2007)\citenamefont
  {Fennell}, \citenamefont {Bramwell}, \citenamefont {McMorrow}, \citenamefont
  {Manuel},\ and\ \citenamefont {Wildes}}]{fennell07}%
  \BibitemOpen
  \bibfield  {author} {\bibinfo {author} {\bibfnamefont {T.}~\bibnamefont
  {Fennell}}, \bibinfo {author} {\bibfnamefont {S.~T.}\ \bibnamefont
  {Bramwell}}, \bibinfo {author} {\bibfnamefont {D.~F.}\ \bibnamefont
  {McMorrow}}, \bibinfo {author} {\bibfnamefont {P.}~\bibnamefont {Manuel}}, \
  and\ \bibinfo {author} {\bibfnamefont {A.~R.}\ \bibnamefont {Wildes}},\
  }\href@noop {} {\bibfield  {journal} {\bibinfo  {journal} {Nat. Phys.}\
  }\textbf {\bibinfo {volume} {3}},\ \bibinfo {pages} {566} (\bibinfo {year}
  {2007})}\BibitemShut {NoStop}%
\bibitem [{\citenamefont {Ruff}\ \emph {et~al.}(2005)\citenamefont {Ruff},
  \citenamefont {Melko},\ and\ \citenamefont {Gingras}}]{Ruff2005}%
  \BibitemOpen
  \bibfield  {author} {\bibinfo {author} {\bibfnamefont {J.~P.~C.}\
  \bibnamefont {Ruff}}, \bibinfo {author} {\bibfnamefont {R.~G.}\ \bibnamefont
  {Melko}}, \ and\ \bibinfo {author} {\bibfnamefont {M.~J.~P.}\ \bibnamefont
  {Gingras}},\ }\href@noop {} {\bibfield  {journal} {\bibinfo  {journal} {Phys.
  Rev. Lett.}\ }\textbf {\bibinfo {volume} {95}},\ \bibinfo {pages} {097202}
  (\bibinfo {year} {2005})}\BibitemShut {NoStop}%
\bibitem [{\citenamefont {Sato}\ \emph {et~al.}(2006)\citenamefont {Sato},
  \citenamefont {Matsuhira}, \citenamefont {Tayama}, \citenamefont {Hiroi},
  \citenamefont {Takagi},\ and\ \citenamefont {Sakakibara}}]{Sato1}%
  \BibitemOpen
  \bibfield  {author} {\bibinfo {author} {\bibfnamefont {H.}~\bibnamefont
  {Sato}}, \bibinfo {author} {\bibfnamefont {K.}~\bibnamefont {Matsuhira}},
  \bibinfo {author} {\bibfnamefont {T.}~\bibnamefont {Tayama}}, \bibinfo
  {author} {\bibfnamefont {Z.}~\bibnamefont {Hiroi}}, \bibinfo {author}
  {\bibfnamefont {S.}~\bibnamefont {Takagi}}, \ and\ \bibinfo {author}
  {\bibfnamefont {T.}~\bibnamefont {Sakakibara}},\ }\href@noop {} {\bibfield
  {journal} {\bibinfo  {journal} {Journal of Physics: Condensed Matter}\
  }\textbf {\bibinfo {volume} {18}},\ \bibinfo {pages} {L297} (\bibinfo {year}
  {2006})}\BibitemShut {NoStop}%
\bibitem [{\citenamefont {Sato}\ \emph {et~al.}(2007)\citenamefont {Sato},
  \citenamefont {Matsuhira}, \citenamefont {Sakakibara}, \citenamefont
  {Tayama}, \citenamefont {Hiroi},\ and\ \citenamefont {Takagi}}]{Sato2}%
  \BibitemOpen
  \bibfield  {author} {\bibinfo {author} {\bibfnamefont {H.}~\bibnamefont
  {Sato}}, \bibinfo {author} {\bibfnamefont {K.}~\bibnamefont {Matsuhira}},
  \bibinfo {author} {\bibfnamefont {T.}~\bibnamefont {Sakakibara}}, \bibinfo
  {author} {\bibfnamefont {T.}~\bibnamefont {Tayama}}, \bibinfo {author}
  {\bibfnamefont {Z.}~\bibnamefont {Hiroi}}, \ and\ \bibinfo {author}
  {\bibfnamefont {S.}~\bibnamefont {Takagi}},\ }\href@noop {} {\bibfield
  {journal} {\bibinfo  {journal} {Journal of Physics: Condensed Matter}\
  }\textbf {\bibinfo {volume} {19}},\ \bibinfo {pages} {145272} (\bibinfo
  {year} {2007})}\BibitemShut {NoStop}%
\bibitem [{\citenamefont {Schiffer}\ \emph {et~al.}(1994)\citenamefont
  {Schiffer}, \citenamefont {Ramirez}, \citenamefont {Huse},\ and\
  \citenamefont {Valentino}}]{Schiffer1994}%
  \BibitemOpen
  \bibfield  {author} {\bibinfo {author} {\bibfnamefont {P.}~\bibnamefont
  {Schiffer}}, \bibinfo {author} {\bibfnamefont {A.~P.}\ \bibnamefont
  {Ramirez}}, \bibinfo {author} {\bibfnamefont {D.~A.}\ \bibnamefont {Huse}}, \
  and\ \bibinfo {author} {\bibfnamefont {A.~J.}\ \bibnamefont {Valentino}},\
  }\href@noop {} {\bibfield  {journal} {\bibinfo  {journal} {Phys. Rev. Lett.}\
  }\textbf {\bibinfo {volume} {73}},\ \bibinfo {pages} {2500} (\bibinfo {year}
  {1994})}\BibitemShut {NoStop}%
\bibitem [{\citenamefont {Quilliam}\ \emph {et~al.}(2011)\citenamefont
  {Quilliam}, \citenamefont {Yaraskavitch}, \citenamefont {Dabkowska},
  \citenamefont {Gaulin},\ and\ \citenamefont {Kycia}}]{Quilliam2011}%
  \BibitemOpen
  \bibfield  {author} {\bibinfo {author} {\bibfnamefont {J.~A.}\ \bibnamefont
  {Quilliam}}, \bibinfo {author} {\bibfnamefont {L.~R.}\ \bibnamefont
  {Yaraskavitch}}, \bibinfo {author} {\bibfnamefont {H.~A.}\ \bibnamefont
  {Dabkowska}}, \bibinfo {author} {\bibfnamefont {B.~D.}\ \bibnamefont
  {Gaulin}}, \ and\ \bibinfo {author} {\bibfnamefont {J.~B.}\ \bibnamefont
  {Kycia}},\ }\href@noop {} {\bibfield  {journal} {\bibinfo  {journal} {Phys.
  Rev. B}\ }\textbf {\bibinfo {volume} {83}},\ \bibinfo {pages} {094424}
  (\bibinfo {year} {2011})}\BibitemShut {NoStop}%
\bibitem [{\citenamefont {Ruminy}\ \emph {et~al.}(2016)\citenamefont {Ruminy},
  \citenamefont {Groitl}, \citenamefont {Keller}, ,\ and\ \citenamefont
  {Fennell}}]{ruminy16}%
  \BibitemOpen
  \bibfield  {author} {\bibinfo {author} {\bibfnamefont {M.}~\bibnamefont
  {Ruminy}}, \bibinfo {author} {\bibfnamefont {F.}~\bibnamefont {Groitl}},
  \bibinfo {author} {\bibfnamefont {T.}~\bibnamefont {Keller}}, , \ and\
  \bibinfo {author} {\bibfnamefont {T.}~\bibnamefont {Fennell}},\ }\href@noop
  {} {\bibfield  {journal} {\bibinfo  {journal} {Phys. Rev. B}\ }\textbf
  {\bibinfo {volume} {94}},\ \bibinfo {pages} {174406} (\bibinfo {year}
  {2016})}\BibitemShut {NoStop}%
\bibitem [{\citenamefont {Yavors'kii}\ \emph {et~al.}(2008)\citenamefont
  {Yavors'kii}, \citenamefont {Fennell}, \citenamefont {Gingras},\ and\
  \citenamefont {Bramwell}}]{Yavo08}%
  \BibitemOpen
  \bibfield  {author} {\bibinfo {author} {\bibfnamefont {T.}~\bibnamefont
  {Yavors'kii}}, \bibinfo {author} {\bibfnamefont {T.}~\bibnamefont {Fennell}},
  \bibinfo {author} {\bibfnamefont {M.~J.~P.}\ \bibnamefont {Gingras}}, \ and\
  \bibinfo {author} {\bibfnamefont {S.~T.}\ \bibnamefont {Bramwell}},\
  }\href@noop {} {\bibfield  {journal} {\bibinfo  {journal} {Phys. Rev. Lett.}\
  }\textbf {\bibinfo {volume} {101}},\ \bibinfo {pages} {037204} (\bibinfo
  {year} {2008})}\BibitemShut {NoStop}%
\bibitem [{\citenamefont {Henelius}\ \emph {et~al.}(2016)\citenamefont
  {Henelius}, \citenamefont {Lin}, \citenamefont {Enjalran}, \citenamefont
  {Hao}, \citenamefont {Rau}, \citenamefont {Altosaar}, \citenamefont
  {Flicker}, \citenamefont {Yavors'kii},\ and\ \citenamefont
  {Gingras}}]{hene16}%
  \BibitemOpen
  \bibfield  {author} {\bibinfo {author} {\bibfnamefont {P.}~\bibnamefont
  {Henelius}}, \bibinfo {author} {\bibfnamefont {T.}~\bibnamefont {Lin}},
  \bibinfo {author} {\bibfnamefont {M.}~\bibnamefont {Enjalran}}, \bibinfo
  {author} {\bibfnamefont {Z.}~\bibnamefont {Hao}}, \bibinfo {author}
  {\bibfnamefont {J.~G.}\ \bibnamefont {Rau}}, \bibinfo {author} {\bibfnamefont
  {J.}~\bibnamefont {Altosaar}}, \bibinfo {author} {\bibfnamefont
  {F.}~\bibnamefont {Flicker}}, \bibinfo {author} {\bibfnamefont
  {T.}~\bibnamefont {Yavors'kii}}, \ and\ \bibinfo {author} {\bibfnamefont
  {M.~J.~P.}\ \bibnamefont {Gingras}},\ }\href@noop {} {\bibfield  {journal}
  {\bibinfo  {journal} {Phys. Rev. B}\ }\textbf {\bibinfo {volume} {93}},\
  \bibinfo {pages} {024402} (\bibinfo {year} {2016})}\BibitemShut {NoStop}%
\bibitem [{\citenamefont {Prabhakaran}\ and\ \citenamefont
  {Boothroyd}(2011)}]{Prabhak}%
  \BibitemOpen
  \bibfield  {author} {\bibinfo {author} {\bibfnamefont {D.}~\bibnamefont
  {Prabhakaran}}\ and\ \bibinfo {author} {\bibfnamefont {A.}~\bibnamefont
  {Boothroyd}},\ }\href@noop {} {\bibfield  {journal} {\bibinfo  {journal}
  {Crys. Growth}\ }\textbf {\bibinfo {volume} {318}},\ \bibinfo {pages} {1053}
  (\bibinfo {year} {2011})}\BibitemShut {NoStop}%
\bibitem [{epi()}]{epi}%
  \BibitemOpen
  \href@noop {} {}\bibinfo {note} {SurfaceNet GmbH,
  http://www.surfacenet.de}\BibitemShut {NoStop}%
\bibitem [{\citenamefont {Griffiths}(1999)}]{griffiths}%
  \BibitemOpen
  \bibfield  {author} {\bibinfo {author} {\bibfnamefont {D.~J.}\ \bibnamefont
  {Griffiths}},\ }\href@noop {} {\emph {\bibinfo {title} {Introduction to
  Electrodynamics}}},\ \bibinfo {edition} {3rd}\ ed.\ (\bibinfo  {publisher}
  {Prentice-Hall},\ \bibinfo {year} {1999})\BibitemShut {NoStop}%
\bibitem [{\citenamefont {Griffiths}(1982)}]{GriffPaper82}%
  \BibitemOpen
  \bibfield  {author} {\bibinfo {author} {\bibfnamefont {D.~J.}\ \bibnamefont
  {Griffiths}},\ }\href@noop {} {\bibfield  {journal} {\bibinfo  {journal}
  {American Journal of Physics}\ }\textbf {\bibinfo {volume} {50}},\ \bibinfo
  {pages} {698} (\bibinfo {year} {1982})}\BibitemShut {NoStop}%
\bibitem [{\citenamefont {Jackson}(1975)}]{jack}%
  \BibitemOpen
  \bibfield  {author} {\bibinfo {author} {\bibfnamefont {J.~D.}\ \bibnamefont
  {Jackson}},\ }\href@noop {} {\emph {\bibinfo {title} {Classical
  Electrodynamics}}},\ \bibinfo {edition} {2nd}\ ed.\ (\bibinfo  {publisher}
  {John Wiley \& Sons},\ \bibinfo {year} {1975})\BibitemShut {NoStop}%
\bibitem [{\citenamefont {Ewald}(1921)}]{ewal21}%
  \BibitemOpen
  \bibfield  {author} {\bibinfo {author} {\bibfnamefont {P.~P.}\ \bibnamefont
  {Ewald}},\ }\href@noop {} {\bibfield  {journal} {\bibinfo  {journal} {Ann.
  Phys.}\ }\textbf {\bibinfo {volume} {369}},\ \bibinfo {pages} {253 }
  (\bibinfo {year} {1921})}\BibitemShut {NoStop}%
\bibitem [{\citenamefont {Gingras}\ and\ \citenamefont
  {Henelius}(2011)}]{gingras11}%
  \BibitemOpen
  \bibfield  {author} {\bibinfo {author} {\bibfnamefont {M.~J.~P.}\
  \bibnamefont {Gingras}}\ and\ \bibinfo {author} {\bibfnamefont
  {P.}~\bibnamefont {Henelius}},\ }\href@noop {} {\bibfield  {journal}
  {\bibinfo  {journal} {J. Phys.: Conf. Ser.}\ }\textbf {\bibinfo {volume}
  {320}},\ \bibinfo {pages} {012001} (\bibinfo {year} {2011})}\BibitemShut
  {NoStop}%
\bibitem [{\citenamefont {Melko}\ and\ \citenamefont
  {Gingras}(2004)}]{melko04}%
  \BibitemOpen
  \bibfield  {author} {\bibinfo {author} {\bibfnamefont {R.~G.}\ \bibnamefont
  {Melko}}\ and\ \bibinfo {author} {\bibfnamefont {M.~J.~P.}\ \bibnamefont
  {Gingras}},\ }\href@noop {} {\bibfield  {journal} {\bibinfo  {journal} {J.
  Phys.: Condens. Matter}\ }\textbf {\bibinfo {volume} {16}},\ \bibinfo {pages}
  {R1277} (\bibinfo {year} {2004})}\BibitemShut {NoStop}%
\bibitem [{pro()}]{process}%
  \BibitemOpen
  \href@noop {} {}\bibinfo {note} {Intel Xeon E5-2698v3 processors with Cray
  Aries interconnect between nodes}\BibitemShut {NoStop}%
\bibitem [{\citenamefont {Levy}(1968)}]{levy68}%
  \BibitemOpen
  \bibfield  {author} {\bibinfo {author} {\bibfnamefont {P.}~\bibnamefont
  {Levy}},\ }\href@noop {} {\bibfield  {journal} {\bibinfo  {journal} {Phys.
  Rev.}\ }\textbf {\bibinfo {volume} {170}},\ \bibinfo {pages} {595} (\bibinfo
  {year} {1968})}\BibitemShut {NoStop}%
\bibitem [{\citenamefont {Ruelle}(1963)}]{ruelle63}%
  \BibitemOpen
  \bibfield  {author} {\bibinfo {author} {\bibfnamefont {D.}~\bibnamefont
  {Ruelle}},\ }\href@noop {} {\bibfield  {journal} {\bibinfo  {journal} {Helv.
  Phys. Acta}\ }\textbf {\bibinfo {volume} {36}},\ \bibinfo {pages} {183}
  (\bibinfo {year} {1963})}\BibitemShut {NoStop}%
\bibitem [{\citenamefont {Campa}\ \emph {et~al.}(2014)\citenamefont {Campa},
  \citenamefont {Dauxios}, \citenamefont {Fanelli},\ and\ \citenamefont
  {Ruffo}}]{campa14}%
  \BibitemOpen
  \bibfield  {author} {\bibinfo {author} {\bibfnamefont {A.}~\bibnamefont
  {Campa}}, \bibinfo {author} {\bibfnamefont {T.}~\bibnamefont {Dauxios}},
  \bibinfo {author} {\bibfnamefont {D.}~\bibnamefont {Fanelli}}, \ and\
  \bibinfo {author} {\bibfnamefont {S.}~\bibnamefont {Ruffo}},\ }\href@noop {}
  {\emph {\bibinfo {title} {Physics of long-range interacting systems}}},\
  \bibinfo {edition} {1st}\ ed.\ (\bibinfo  {publisher} {Oxford University
  Press},\ \bibinfo {year} {2014})\BibitemShut {NoStop}%
\bibitem [{\citenamefont {Gingras}(2011)}]{gingras11a}%
  \BibitemOpen
  \bibfield  {author} {\bibinfo {author} {\bibfnamefont {M.~J.~P.}\
  \bibnamefont {Gingras}},\ }\enquote {\bibinfo {title} {Introduction to
  frustrated magnetism},}\ \ (\bibinfo  {publisher} {Springer (Ed. Lacroix,
  Mendels \& Mila)},\ \bibinfo {year} {2011})\ Chap.\ \bibinfo {chapter} {Spin
  Ice}\BibitemShut {NoStop}%
\bibitem [{\citenamefont {S\'anchez}\ \emph {et~al.}(2017)\citenamefont
  {S\'anchez}, \citenamefont {Mendoza~Z\'elis}, \citenamefont {Arciniegas},
  \citenamefont {Pasquevich},\ and\ \citenamefont {Fern\'andez~van
  Raap}}]{sanchez17}%
  \BibitemOpen
  \bibfield  {author} {\bibinfo {author} {\bibfnamefont {F.~H.}\ \bibnamefont
  {S\'anchez}}, \bibinfo {author} {\bibfnamefont {P.}~\bibnamefont
  {Mendoza~Z\'elis}}, \bibinfo {author} {\bibfnamefont {M.~L.}\ \bibnamefont
  {Arciniegas}}, \bibinfo {author} {\bibfnamefont {G.~A.}\ \bibnamefont
  {Pasquevich}}, \ and\ \bibinfo {author} {\bibfnamefont {M.~B.}\ \bibnamefont
  {Fern\'andez~van Raap}},\ }\href@noop {} {\bibfield  {journal} {\bibinfo
  {journal} {Phys. Rev. B}\ }\textbf {\bibinfo {volume} {95}},\ \bibinfo
  {pages} {134421} (\bibinfo {year} {2017})}\BibitemShut {NoStop}%
\bibitem [{\citenamefont {Rhodes}\ and\ \citenamefont
  {Rowlands}(1954)}]{rhodes54}%
  \BibitemOpen
  \bibfield  {author} {\bibinfo {author} {\bibfnamefont {P.}~\bibnamefont
  {Rhodes}}\ and\ \bibinfo {author} {\bibfnamefont {G.}~\bibnamefont
  {Rowlands}},\ }\href@noop {} {\bibfield  {journal} {\bibinfo  {journal}
  {Proc. Leeds Phil. Lit. Soc.}\ }\textbf {\bibinfo {volume} {6}},\ \bibinfo
  {pages} {191} (\bibinfo {year} {1954})}\BibitemShut {NoStop}%
\bibitem [{\citenamefont {Bramwell}\ \emph {et~al.}(2000)\citenamefont
  {Bramwell}, \citenamefont {Field}, \citenamefont {Harris},\ and\
  \citenamefont {Parkin}}]{Bramwell00}%
  \BibitemOpen
  \bibfield  {author} {\bibinfo {author} {\bibfnamefont {S.~T.}\ \bibnamefont
  {Bramwell}}, \bibinfo {author} {\bibfnamefont {M.~N.}\ \bibnamefont {Field}},
  \bibinfo {author} {\bibfnamefont {M.~J.}\ \bibnamefont {Harris}}, \ and\
  \bibinfo {author} {\bibfnamefont {I.~P.}\ \bibnamefont {Parkin}},\ }\href
  {http://stacks.iop.org/0953-8984/12/i=4/a=308} {\bibfield  {journal}
  {\bibinfo  {journal} {Journal of Physics: Condensed Matter}\ }\textbf
  {\bibinfo {volume} {12}},\ \bibinfo {pages} {483} (\bibinfo {year}
  {2000})}\BibitemShut {NoStop}%
\bibitem [{mpi()}]{mpi}%
  \BibitemOpen
  \href@noop {} {}\bibinfo {note} {Open MPI v2.0.0 documentation.
  https://www.open-mpi.org/doc/v2.0/}\BibitemShut {NoStop}%
\bibitem [{\citenamefont {Kunkel}\ \emph {et~al.}(2016)\citenamefont {Kunkel},
  \citenamefont {Balaji},\ and\ \citenamefont {Dongarra}}]{scaling}%
  \BibitemOpen
  \bibfield  {author} {\bibinfo {author} {\bibfnamefont {J.}~\bibnamefont
  {Kunkel}}, \bibinfo {author} {\bibfnamefont {P.}~\bibnamefont {Balaji}}, \
  and\ \bibinfo {author} {\bibfnamefont {J.}~\bibnamefont {Dongarra}},\ }\href
  {https://books.google.se/books?id=kRJkDAAAQBAJ} {\enquote {\bibinfo {title}
  {High performance computing: 31st international conference},}\ } (\bibinfo
  {year} {2016}),\ \bibinfo {note} {{I}SC High Performance 2016, Frankfurt,
  Germany, June 19-23, 2016, Proceedings}\BibitemShut {NoStop}%
\bibitem [{\citenamefont {Hastings}(1970)}]{Hast}%
  \BibitemOpen
  \bibfield  {author} {\bibinfo {author} {\bibfnamefont {W.~K.}\ \bibnamefont
  {Hastings}},\ }\href@noop {} {\bibfield  {journal} {\bibinfo  {journal}
  {Biometrika}\ }\textbf {\bibinfo {volume} {57}},\ \bibinfo {pages} {97}
  (\bibinfo {year} {1970})}\BibitemShut {NoStop}%
\bibitem [{\citenamefont {Melko}\ \emph {et~al.}(2001)\citenamefont {Melko},
  \citenamefont {den Hertog},\ and\ \citenamefont {Gingras}}]{loopalgo}%
  \BibitemOpen
  \bibfield  {author} {\bibinfo {author} {\bibfnamefont {R.~G.}\ \bibnamefont
  {Melko}}, \bibinfo {author} {\bibfnamefont {B.~C.}\ \bibnamefont {den
  Hertog}}, \ and\ \bibinfo {author} {\bibfnamefont {M.~J.~P.}\ \bibnamefont
  {Gingras}},\ }\href {\doibase 10.1103/PhysRevLett.87.067203} {\bibfield
  {journal} {\bibinfo  {journal} {Phys. Rev. Lett.}\ }\textbf {\bibinfo
  {volume} {87}},\ \bibinfo {pages} {067203} (\bibinfo {year}
  {2001})}\BibitemShut {NoStop}%
\bibitem [{\citenamefont {den Hertog}\ and\ \citenamefont
  {Gingras}(2000)}]{Hertog2000}%
  \BibitemOpen
  \bibfield  {author} {\bibinfo {author} {\bibfnamefont {B.~C.}\ \bibnamefont
  {den Hertog}}\ and\ \bibinfo {author} {\bibfnamefont {M.~J.~P.}\ \bibnamefont
  {Gingras}},\ }\href@noop {} {\bibfield  {journal} {\bibinfo  {journal} {Phys.
  Rev. Lett.}\ }\textbf {\bibinfo {volume} {84}},\ \bibinfo {pages} {3430}
  (\bibinfo {year} {2000})}\BibitemShut {NoStop}%
\end{thebibliography}
\end{document}